\documentstyle[psfig,epsfig]{mn}

\newcommand{\HI}{H\,{\sc i}}
\newcommand{\HII}{H\,{\sc ii}}
\newcommand{\NII}{[N\,{\sc ii}]}
\newcommand{\HeII}{He\,{\sc ii}}
\newcommand{\Ha}{H$\alpha$}
\newcommand{\Hb}{H$\beta$}
\newcommand{\kms}{~km\,s$^{-1}$}
\newcommand{\kkms}{km\,s$^{-1}$}

\newcommand{\vopt}{$v_{\rm opt}$}

\newcommand{\vsys}{$v_{\rm sys}$}
\newcommand{\vlg}{$v_{\rm LG}$}

\newcommand{\vrot}{$v_{\rm rot}$}
\newcommand{\FHI}{$F_{\rm HI}$}

\newcommand{\LB}{$L_{\rm B}$}

\newcommand{\MHI}{$M_{\rm HI}$}

\newcommand{\Msun}{~M$_{\odot}$}
\newcommand{\Lsun}{~L$_{\odot}$}
\newcommand{\Zsun}{~Z$_{\odot}$}

\newcommand{\AB}{A$_{\rm B}$}

\newcommand{\Mdyn}{$M_{\rm dyn}$}

\newcommand{\sfrha}{$SFR_{\rm H\alpha}$}
\newcommand{\sfrmir}{$SFR_{\rm H\alpha+24\mu m}$}
\newcommand{\sfrfir}{$SFR_{\rm FIR}$}

\newcommand{\sfrghz}{$SFR_{\rm 20cm}$}
\newcommand{\sfrfuv}{$SFR_{\rm FUV}$}
\newcommand{\Moy}{~M$_{\odot}$\,yr$^{-1}$}
\newcommand{\MMoy}{M$_{\odot}$\,yr$^{-1}$}

\title[Gas Dynamics and Star Formation in the Galaxy Pair NGC~1512/1510]
      {Gas Dynamics and Star Formation \\ 
          in the Galaxy Pair NGC~1512/1510\thanks{
       The observations were obtained with the Australia Telescope which is 
       funded by the Commonwealth of Australia for operations as a National 
       Facility managed by CSIRO.}}

\author[Koribalski \& L\'opez-S\'anchez]
       {B\"arbel S. Koribalski \& \'Angel R. L\'opez-S\'anchez \\
        Australia Telescope National Facility, CSIRO, 
            P.O. Box 76, Epping, NSW 1710, Australia 
}

\date{Received date: 2008, Sep 27; accepted date: 2009, Aug 24}

\begin{document}

\maketitle

\begin{abstract}
Here we present \HI\ line and 20-cm radio continuum data of the nearby galaxy 
pair NGC~1512/1510 as obtained with the Australia Telescope Compact Array 
(ATCA). These are complemented by GALEX $UV$-, SINGG \Ha- and Spitzer 
mid-infrared images, allowing us to compare the distribution and kinematics 
of the neutral atomic gas with the locations and ages of the stellar clusters 
within the system.
 
For the barred, double-ring galaxy NGC~1512 we find a very large \HI\ disk,
$\sim$4 $\times$ its optical diameter, with two pronounced spiral/tidal arms.
Both its gas distribution and the distribution of the star-forming regions are
affected by gravitational interaction with the neighbouring blue compact dwarf 
galaxy NGC~1510. While the inner disk of NGC~1512 shows quite regular rotation,
deviations are visible along the outer arms and at the position of NGC~1510. 
From the \HI\ rotation curve of NGC~1512 we estimate a dynamical mass of \Mdyn\ 
$\ga 3 \times 10^{11}$\Msun, compared to an \HI\ mass of \MHI\ = $5.7 \times 
10^9$\Msun\ ($\sim$2\% \Mdyn).

The two most distant \HI\ clumps, at radii of $\sim$80 kpc, show signs of star 
formation and are likely {\em tidal dwarf galaxies} (TDGs). Both lie along an
extrapolation of the eastern-most \HI\ arm, with the most compact \HI\ cloud 
located at the tip of the arm.
 
The 20-cm radio continuum map indicates extended star formation activity not 
only in the central regions of both galaxies but also in between them. Star 
formation (SF) in the outer disk of NGC~1512 is revealed by deep optical- and 
two-color ultraviolet images. Using the latter we determine the properties 
of $\ga$200 stellar clusters and explore their correlation with dense \HI\ 
clumps in the even larger 2X\HI\ disk. Outside the inner star-forming ring of 
NGC~1512, which must contain a large reservoir of molecular gas, \HI\ turns 
out to be an excellent tracer of SF activity. 

The multi-wavelength analysis of the NGC~1512/1510 system, which is probably 
in the first stages of a minor merger having started $\sim$400~Myr ago, links
stellar and gaseous galaxy properties on scales from one to 100 kpc. 
\end{abstract}

 
\section{Introduction} 

The Local Volume (LV), generally considered as the sphere of radius 10 Mpc 
centred on the Local Group, contains more than 500 galaxies. For the majority 
of these galaxies reliable distances are currently available (Karachentsev 
et al. 2004, 2008). Independent distances, such as those obtained from the 
luminosity of Cepheids, the tip of the red giant branch (TRGB), and surface 
brightness fluctuations (SBF) are an essential ingredient, together with 
accurate velocities and detailed multi-wavelength studies of each LV galaxy, 
for the assembly of a dynamic 3D view of the Local Universe. This, in turn, 
leads to a better understanding of the local flow field, the local mass 
density and the local star-formation density. 
Interferometric \HI\ measurements, in particular, provide insight into the 
overall matter distribution (baryonic and non-baryonic) in the Local Volume.

\begin{figure*} 
\label{fig:figure1}  
\psfig{file=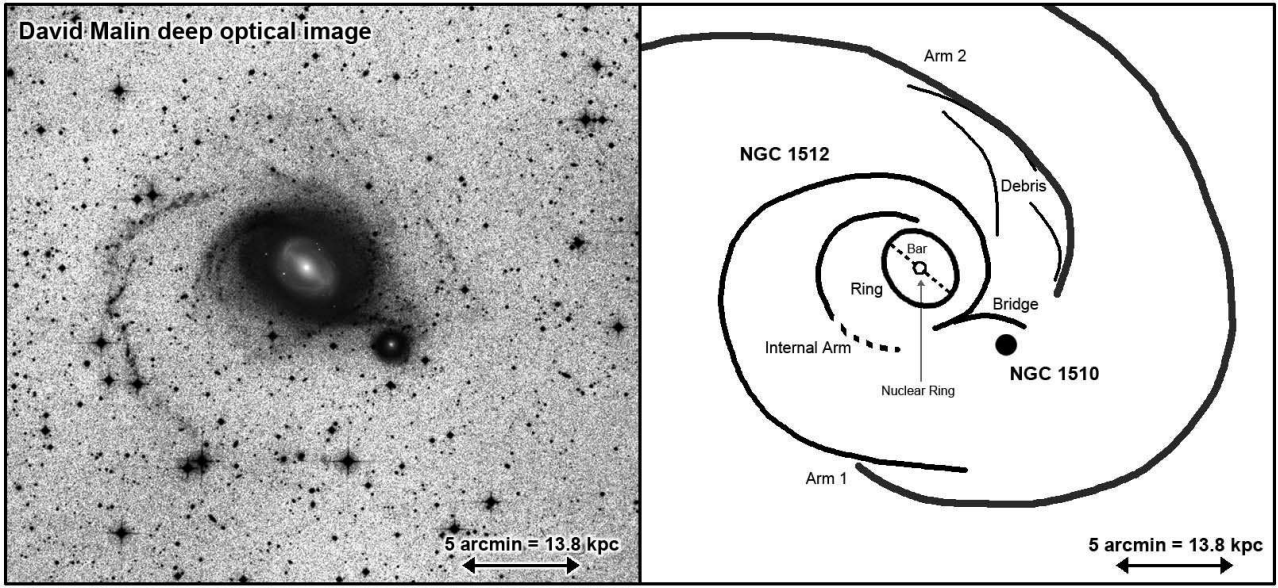,width=17cm,angle=0}
\caption{({\bf Left}) Deep optical image of the galaxy pair NGC~1512/1510 
   obtained by David Malin (priv. com.) from combined UK Schmidt Telescope 
   plates; it has been saturated to emphasise the faintest stellar structures 
   of the system, in particular the prominent eastern arm and the bridge 
   between NGC~1512 and NGC~1510. The grey scale is logarithmic; the displayed
   field of view is $27\arcmin \times 25\arcmin$. A non-saturated $R$-band 
   image of the pair, obtained as part of the SINGS project (Kennicutt et 
   al. 2003), is overlaid onto the central region. ({\bf Right}) Sketch of 
   the stellar and \HI\ structure of the NGC~1512/1510 system; see text for 
   details.}
\end{figure*}

The galaxy pair NGC~1512/1510 is located in the outskirts of the Local Volume 
and its study forms part of the `Local Volume \HI\ Survey' (LVHIS; Koribalski 
et al. 2008). Since no TRGB distance is currently available for NGC~1512, we 
use its Local Group velocity, \vlg\ = 712\kms, to compute a Hubble distance 
of $\sim$9.5 Mpc. --- LVHIS is a large project\footnote{LVHIS project:~~~
   www.atnf.csiro.au/research/LVHIS} that aims to provide detailed \HI\ 
distributions, velocity fields and star formation rates for a complete sample 
of nearby, gas-rich galaxies. With the Australia Telescope Compact Array 
(ATCA), we observed all LV galaxies that were detected in the \HI\ Parkes 
All-Sky Survey (HIPASS; Barnes et al. 2001, Koribalski et al. 2004) and 
reside south of approx. --30\degr\ declination.

The closest neighbours to the NGC~1512/1510 system are (1) the edge-on spiral 
galaxy NGC~1495 (HIPASS J0358--44), (2) the galaxy pair NGC~1487 (HIPASS 
J0355--42) and (3) the galaxy ESO249-G026 (HIPASS J0354--43), all located at 
projected distances of more than 1\fdg5. Within 3\degr\ ($\sim$0.5 Mpc) we 
find 15 neighbours, suggesting that the NGC~1512/1510 system is part of a 
loose (spiral) galaxy group (LGG\,108; Garcia 1993).

The barred galaxy NGC~1512 and the blue compact dwarf (BCD) galaxy NGC~1510 
are an interacting galaxy pair, separated by only $\sim$5\arcmin\ (13.8 kpc).
At the adopted distance of 9.5 Mpc, 1\arcmin\ corresponds to 2.76 kpc. 
Table~1 gives some basic properties of both galaxies.

\begin{table} 
\caption{Basic properties of NGC~1512 and NGC~1510.}
\label{tab:table1} 
\begin{tabular}{lccccc}
\hline
                     & NGC~1512  & NGC~1510                     & Ref. \\
                     & \multicolumn{2}{c}{HIPASS J0403--43}        \\
\hline
center position      & $04^{\rm h}\,03^{\rm m}\,54\fs6$  
                     & $04^{\rm h}\,03^{\rm m}\,32\fs6$         & (1) \\
~$\alpha,\delta$(J2000)& --43\degr\,21\arcmin\,03\arcsec 
                       & --43\degr\,24\arcmin\,01\arcsec \\ 
~$l,b$               & 248\fdg7, --48\fdg2 & 248\fdg8, --48\fdg2& (1) \\
$v_{\rm opt}$ [\kkms]& $896\pm5$ & $989\pm23$                   & (2,1) \\ 
type                 & SB(r)ab   & SA0 pec, BCD                 & (1) \\
optical diameter     & $8\farcm9 \times 5\farcm6$  
                     & $1\farcm3 \times 0\farcm7$               & (1) \\ 
~~~~~~~" (kpc$^2$)   & $24.6 \times 15.5$ &  $3.6 \times 1.9$  \\
inclination          & 51\degr            & 57\degr             & (1) \\
position angle       & 90\degr            & 90\degr             & (1) \\
\AB\ [mag]           & 0.046              & 0.046               & (3) \\ 
$m_{\rm B}$ [mag]    & $11.08\pm0.09$     & $13.47\pm0.11$      & (4) \\
$U-B$                & $0.14\pm0.12$      & $-0.23\pm0.22$      & (4)  \\
$B-V$                & $0.74\pm0.12$      & $0.43\pm0.22$       & (4)  \\
$M_{\rm B}$ [mag]    & $-18.86\pm0.09$    & $-16.47\pm0.11$     & (4) \\
\LB\ [10$^9$\Lsun]   & $5.45\pm0.45$      & $0.60\pm0.08$       & (4) \\
\hline
$v_{\rm HI}$ [\kkms] & \multicolumn{2}{c}{$898\pm3$}        & (5) \\
$v_{\rm LG}$ [\kkms] & \multicolumn{2}{c}{712}              & (5) \\
distance [Mpc]       & \multicolumn{2}{c}{9.5}              & (5) \\ 
$w_{\rm 50}$ [\kkms] & \multicolumn{2}{c}{$234\pm6$}        & (5) \\
$w_{\rm 20}$ [\kkms] & \multicolumn{2}{c}{$270\pm9$}        & (5) \\
\FHI\ [Jy\kms]       & \multicolumn{2}{c}{$259.3 \pm 17.4$} & (5) \\
\MHI\ [10$^9$\Msun]  & \multicolumn{2}{c}{$5.51 \pm 0.37$}  & (5) \\ 
\hline
\end{tabular}
\flushleft
The $B$-band luminosity, \LB, is calculated from the $B$-band magnitude, 
  $m_{\rm B}$, using a solar $B$-band magnitude of 5.48 mag. Unless 
  otherwise stated, velocities are in the heliocentric frame using the
  optical definition. ---
References: 
  (1) de Vaucouleurs et al. (1991; RC3), 
  (2) Da Costa et al. (1991), 
  (3) Schlegel et al. (1998), 
  (4) using data provided by Gil de Paz et al. (2007a) 
      and correcting for Galactic extinction, \AB,
  (5) Koribalski et al. (2004; HIPASS BGC) and derived properties.
\end{table}

The optical appearances of both galaxies are well described by Hawarden et al. 
(1979). NGC~1512 (IRAS~04022--4329) is a large, strongly barred galaxy with 
two prominent star-forming rings. Its morphological type is generally given 
as SB(r)a or SB(r)b. The companion, NGC~1510 (IRAS~04019--4332), is a much 
smaller, peculiar S0 or lenticular galaxy. Their respective optical diameters 
are $8\farcm9 \times 5\farcm6$ and $1\farcm3 \times 0\farcm7$, i.e. NGC~1512's 
stellar disk is about seven times larger than that of NGC~1510. 

\begin{table*} 
\caption{Summary of the multi-pointing ATCA radio observations 
         of the galaxy pair NGC~1512/1510.}
\label{tab:table2} 
\begin{tabular}{lccccccc}
\hline
ATCA configuration 
     & H168    & 210    & 375     & 750A    & 1.5A     & 6A     & 6B \\
\hline
date & 8-11-05 & 6-7-00 & 23-9-96 & 6-11-96 & 20-10-96 & 5-2-97 & 14-9-96 \\
~~"  &         & 8-7-00 & 24-9-96 & \\
~~"  &         &        & 3-12-96 & \\
time on-source [min.]
     & 450     & 318    & 489     & 653     & 606      & 283    & 633 \\
     &         & 238    & 502     & \\
     &         &        & 305     & \\
primary calibrator & \multicolumn{6}{c}{PKS\,1934--638 (14.95 Jy)} \\
phase calibrator   & \multicolumn{6}{c}{PKS\,0438--436 ( 4.55 Jy)} \\
\hline
\end{tabular}
\end{table*}

Beautiful multi-color HST images of NGC~1512 by Maoz et al. (2001) clearly 
show the structure of the nuclear region ($<$20\arcsec\ $\sim$ 1 kpc): a 
bright nucleus surrounded by a smooth, dusty disk which is enveloped by 
a highly ordered and narrow starburst ring of diameter $16\arcsec \times 
12\arcsec$ with a position angle ($PA$) of $\sim$90\degr. This nuclear ring 
is also evident in the $J-K$ map by Laurikainen et al. (2006) and in the 
Spitzer mid-infrared images obtained as part of the SINGS project (Kennicutt 
et al. 2003). The dust lanes hint at a tight inner spiral structure within 
the nuclear disk. Fabry-Perot \Ha\ observations of NGC~1512 by Buta (1988) 
show that the nuclear ring has a rotational velocity of \vrot\ $\sim$ 
200--220\kms\ (assuming an inclination angle of 35\degr). Beyond the nuclear 
ring, which lies within the bulge ($\la$1\arcmin\  = 2.8 kpc) at the centre 
of the bar, \vrot\ appears to be roughly constant.

\Ha\ images of the inner region ($<$4\arcmin\ $\approx$ 11 kpc) of NGC~1512 
(SINGG project; Meurer et al. 2006) reveal a second star-forming ring of 
approximate diameter $3\arcmin \times 2\arcmin$ at $PA \sim 45\degr$, ie. about
ten times larger than the nuclear starburst ring; its width is $20\arcsec - 
40\arcsec$. This inner ring is composed of dozens of independent \HII\ regions
with typical sizes of $2\arcsec - 5\arcsec$. The bar, which has a length of 
$\sim$3\arcmin\ (8.3~kpc), lies roughly along its major axis. Some enhancement 
of the star formation is seen at both ends of the bar where the spiral arms 
commence.

The optical data presented by Kinman (1978) and Hawarden et al. (1979) 
revealed, for the first time, signs of tidal interaction between NGC~1510 and 
NGC~1512. Sandage \& Bedke (1994) describe NGC~1512 as {\em an almost-normal
SBb(r) where interaction with NGC~1510 distorts the outer thin arm pattern}.
The stellar spiral arms are most prominent in deep optical images (see Fig.~1) 
as well as the GALEX ultraviolet ($UV$) images by Gil de Paz et al. (2007a), 
all of which give a stunning view of the star-forming regions in NGC~1512's 
outer disk. A sketch identifying important stellar and \HI\ features of the 
interacting system NGC~1512/1510 is provided on the right side of Fig.~1.
  
NGC~1510 is a low metallicity ($Z \sim 0.2$\Zsun) BCD galaxy (see also 
Section~4.6). Hawarden et al. (1979) suggested that its emission line spectrum 
and blue colors are the consequence of star formation activity in the material 
--- basically \HI\ gas --- recently ($\sim$300~Myr) accreted from NGC~1512, 
mimicing the properties of a red amorphous dwarf elliptical galaxy. This 
hypothesis is also supported by Eichendorf \& Nieto (1984), who identified 
several low-metallicity star-forming regions in NGC~1510. One of them (the SW 
component) reveals a broad $\lambda$4686 \HeII\ line which is attributed to 
the presence of an important population of Wolf-Rayet (WR) stars in the burst. 
NGC~1510 is therefore classified as Wolf-Rayet galaxy (Conti 1991; Schaerer, 
Contini \& Pindao 1999).

Hawarden et al. (1979) also present remarkable \HI\ data for the galaxy pair.
Their 24-pointing \HI\ map obtained with the 64-m Parkes telescope reveals a 
large neutral hydrogen envelope around NGC~1512, encompassing its neighbour, 
NGC~1510. Koribalski et al. (2004) measure an integrated \HI\ flux density of 
\FHI\ = $259\pm17$ Jy\kms\ for the galaxy pair, named HIPASS J0403--43 in the 
HIPASS Bright Galaxy Catalog (see Table~1). The detected \HI\ emission is 
centered on NGC~1512 and significantly extended with respect to the Parkes 
gridded beam of 15\farcm5. Hawarden et al. (1979) measured \FHI\ = $232\pm20$ 
Jy\kms\ (same as Reif et al. 1982), slightly lower than the HIPASS value.

Here we present high-resolution ATCA \HI\ line and 20-cm radio continuum data 
of the galaxy pair NGC~1512/1510 as well as complimentary GALEX $UV$-, SINGG
\Ha- and Spitzer mid-infrared images.
The paper is organised as follows: in Section~2 we summarise the observations 
and data reduction; in Section~3 we present the \HI\ line and the 20-cm radio 
continuum results, including our discovery of two {\em tidal dwarf galaxy}
candidates. 
The discussion in Section~4 exploits the available multi-wavelength data sets,
comparing the \HI\ gas density with the properties of star-forming regions out 
to radii of 80~kpc. Section~5 contains our conclusions and Section~6 a brief 
outlook towards \HI\ surveys with the Australian SKA Pathfinder (ASKAP).

\begin{figure*}
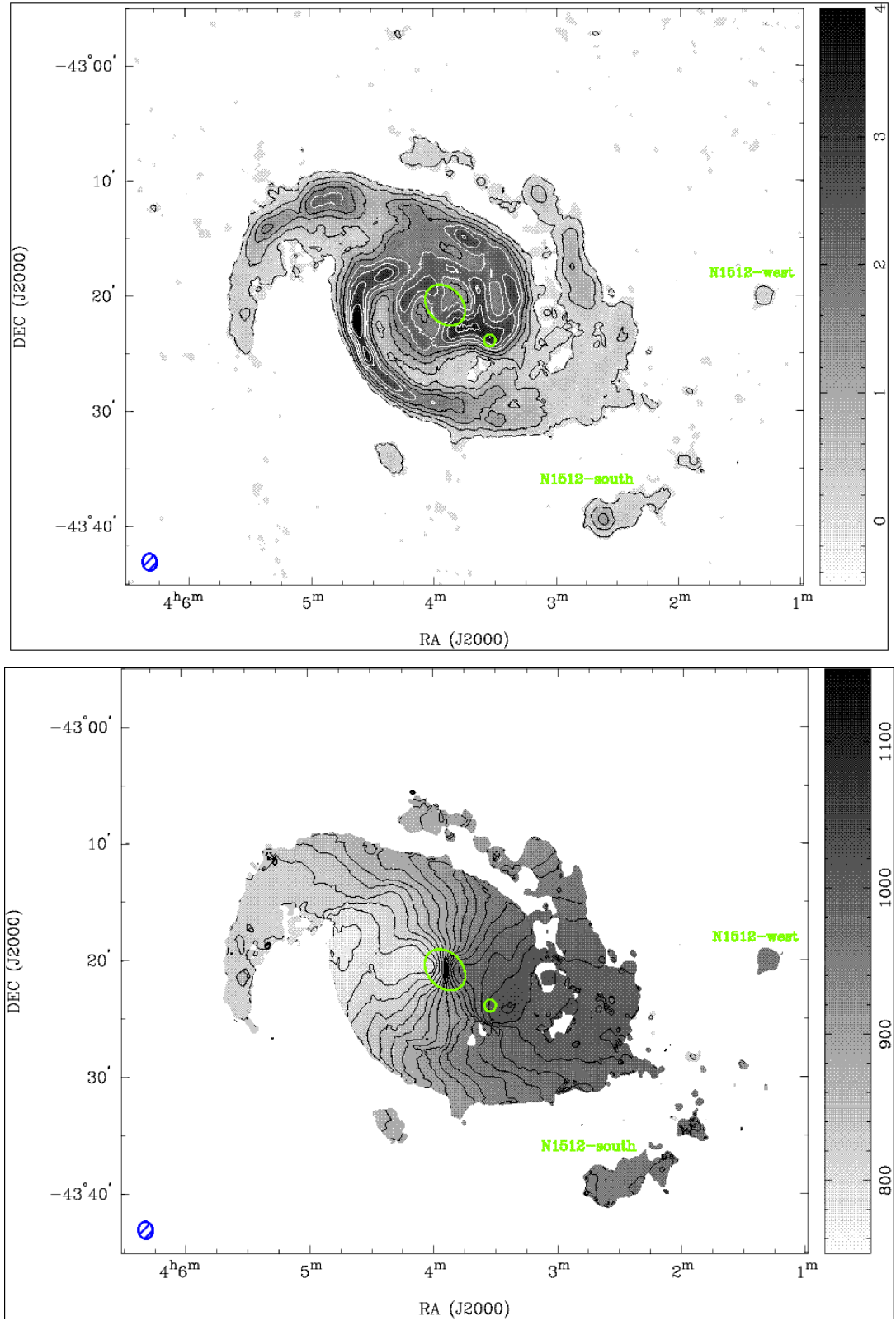
 
\label{fig:figure2}  
\begin{tabular}{c}
 \mbox{\psfig{file=ngc1512.line.na6uv4.mom0.cps,width=10.5cm,angle=-90}} \\
 \mbox{\psfig{file=ngc1512.line.na6uv4.mom1.mask.cps,width=10.5cm,angle=-90}} \\
\end{tabular}
\caption{\HI\ moment maps of the galaxy pair NGC~1512/1510 as obtained from 
   the ATCA using `natural' weighting. 
   Note that the displayed image size is $\sim60\arcmin \times 50\arcmin$, 
   showing about four times the area of Fig.~1.
   {\bf (top)} \HI\ distribution (contour levels: 0.1, 0.5, 1, 1.5, 2, 2.5, 
   3 and 3.5 Jy\,beam$^{-1}$\kms), and {\bf (bottom)} mean, masked \HI\ 
   velocity field (contour levels range from 785 to 1025\kms, in steps of 
   15\kms). The synthesised beam ($88\farcs3 \times 75\farcs5$) is displayed 
   in the bottom left corner of each panel. Assuming the gas fills the beam, 
   0.1 Jy\kms\ corresponds to an \HI\ column density of $1.7 \times 10^{19}$ 
   atoms~cm$^{-2}$. The ellipse (center) and the circle ($\sim5\arcmin$ 
   towards the SW) mark the position and size of the 20-cm radio continuum 
   emission from NGC~1512 and NGC~1510, respectively. --- Note that the \HI\
   emission of the galaxy NGC~1512 extends well beyond the known stellar 
   disk/arms.}
\end{figure*}

\begin{figure*} 
\label{fig:figure3}  
 \mbox{\psfig{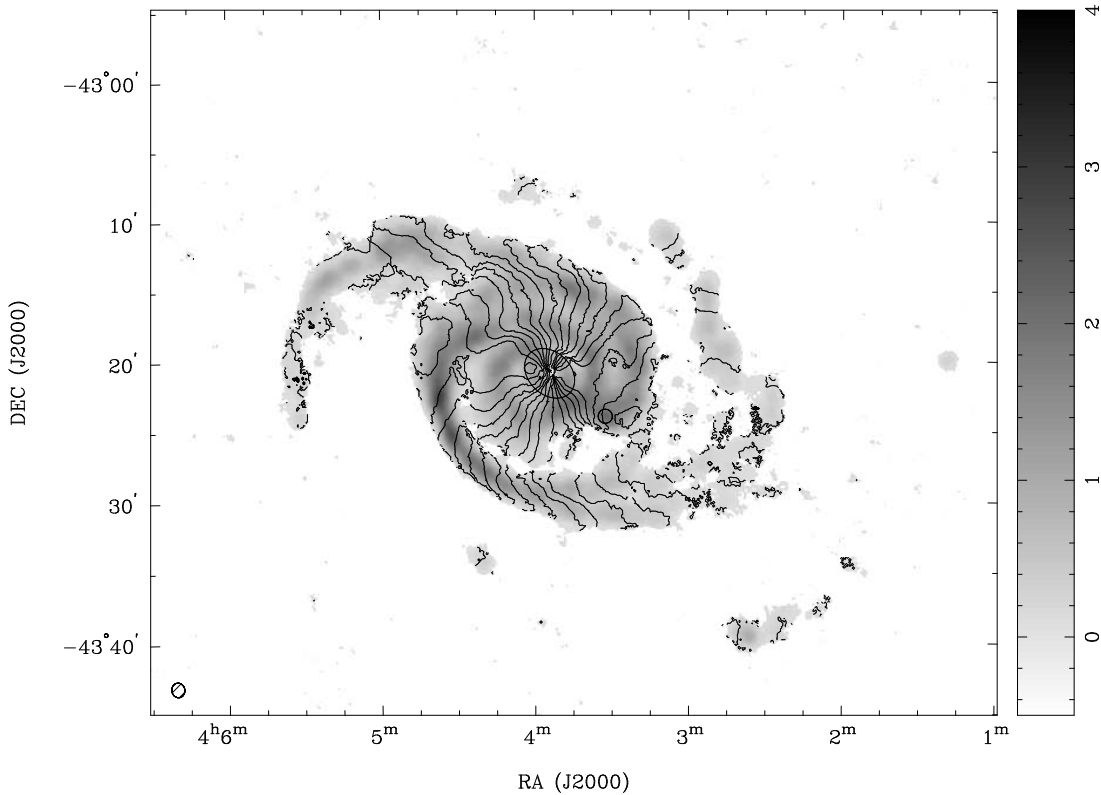}} 
\caption{The mean \HI\ velocity field (contours) of the galaxy pair 
   NGC~1512/1510 overlaid onto the \HI\ distribution (grey scale) at slightly 
   higher angular resolution ($62\farcs1 \times 55\farcs3$) than Fig.~2. 
   The intensity units are Jy\,beam$^{-1}$\kms.}
\end{figure*}

\section{Observations and Data Reduction} 

\HI\ line and 20-cm radio continuum observations of the galaxy pair 
NGC~1512/1510 were obtained with the Australia Telescope Compact Array (ATCA) 
using multiple configurations and four (overlapping) pointings. The observing 
details are given in Table~2.
The first frequency band (IF1) was centered on 1415 MHz with a bandwidth of 
8 MHz, divided into 512 channels. This gives a channel width of 3.3\kms\ and 
a velocity resolution of 4\kms. The ATCA primary beam is 33\farcm6 at 
1415 MHz. The second frequency band (IF2) was centered on 1384 MHz (20-cm) 
with a bandwidth of 128 MHz divided into 32 channels. \\

The ATCA is a radio interferometer consisting of six 22-m dishes, creating
15 baselines in a single configuration, equipped with seven receiver systems 
covering wavelengths from 3-mm to 20-cm. While five antennas (CA01 to CA05) 
are movable along a 3-km long east-west track (and a 214-m long north-south 
spur, allowing us to create hybrid arrays), one antenna (CA06) is fixed at 
a distance of 3-km from the end of the track. By combining data from several 
array configuration (see Table~2) we achieve excellent $uv$-coverage generated 
by over 100 baselines ranging from 30-m to 6-km. Using Fourier transformation,
this allows us to make data cubes and images at a large range of angular 
resolutions (up to 6\arcsec\ at 20-cm) by choosing different weights for 
short, medium and long baselines which in turn are sensitive to different 
structure scales. The weighting of the data affects not only the resolution, 
but also the rms noise and sensitivity to diffuse emission. \\

Data reduction was carried out with the {\sc miriad} software package (Sault,
Teuben \& Wright 1995) using 
standard procedures. After calibration the IF1 data were split into a narrow 
band 20-cm radio continuum and an \HI\ line data set using a first order fit 
to the line-free channels. \HI\ cubes were made using `natural' (na) and 
`robust' (r=0) weighting of the {\em uv}-data in the velocity range covered by 
the \HI\ emission using steps of 10\kms. The longest baselines to the distant 
antenna six (CA06) were excluded when making the low-resolution cubes. 
Broad-band 20-cm radio continuum images were made using `robust' (r=0) and 
`uniform' weighting of the IF2 {\em uv}-data. 
The data were analysed using {\sc miriad}, apart from the rotation curve fit
which was obtained using the {\sc gipsy} software package (van der Hulst et 
al. 1992).

\begin{figure*} 
\label{fig:figure4} 
  \mbox{\psfig{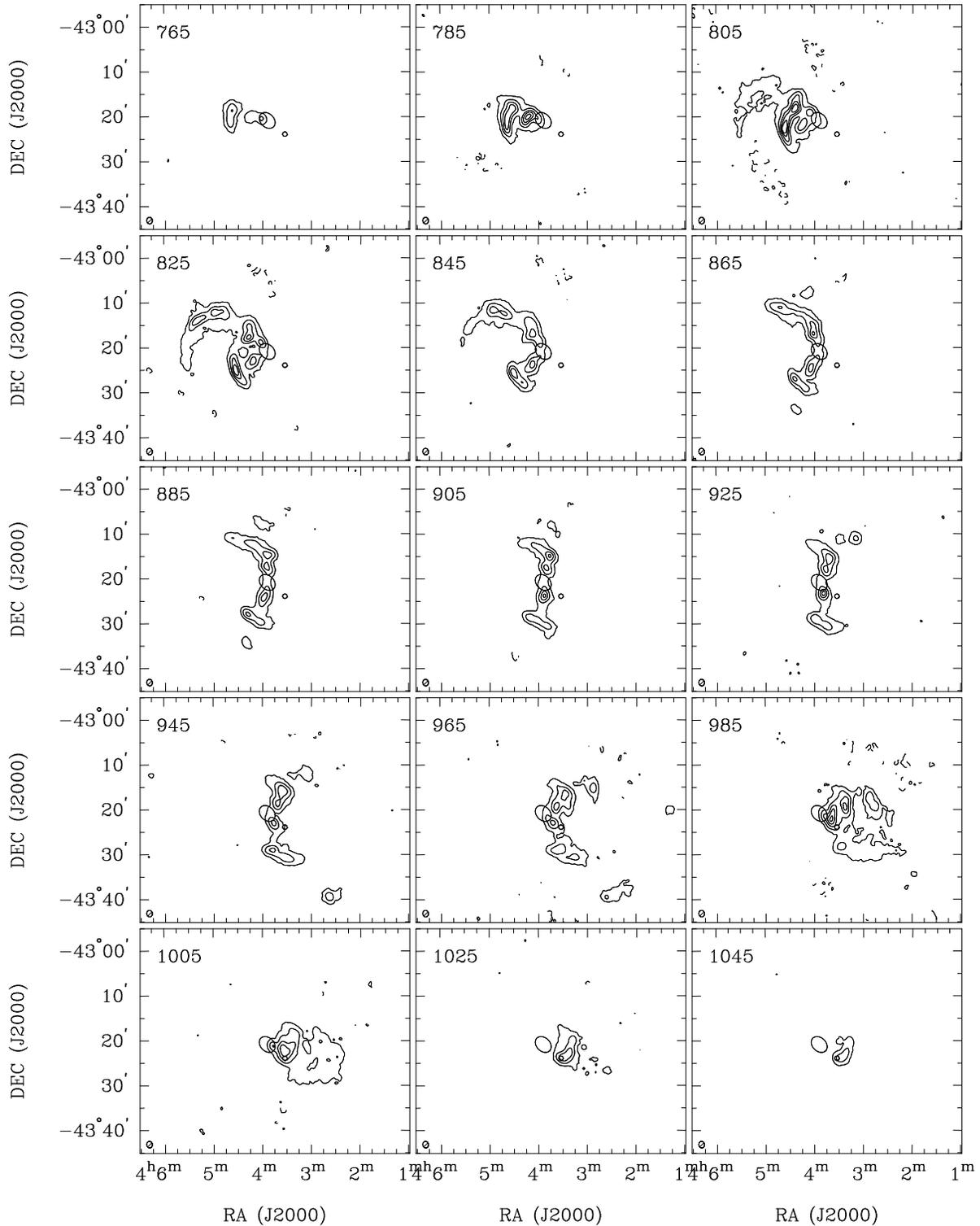}}
\caption{\HI\ channel maps of the galaxy pair NGC~1512/1510 as obtained from 
   the ATCA using `natural' weighting.
   The TDG candidates are visible in the two channels maps at 945 and 965\kms.
   Note that the image size is chosen to be the same as in Figs.~2 \& 3.
   The contour levels are --6, 6, 20, 60, and 80 mJy\,beam$^{-1}$. The two 
   galaxy centres are marked with black (green) ellipses corresponding in size 
   to the detected 20-cm radio continuum emission. For display purposes the
   channel width is 20\kms. The centre velocity of each channel is displayed 
   at the top left and the synthesised beam ($88\farcs3 \times 75\farcs5$) 
   at the bottom left corner of each panel.}
\end{figure*}

\section{Results} 

The NGC~1512/1510 galaxy pair is an impressive system. Our ATCA \HI\ mosaic 
(see Figs.~2--4) shows a very extended gas distribution, spanning a diameter
of $\sim$40\arcmin\ (or 110 kpc). Two prominent spiral arms, which appear to 
wrap around $\sim$1.5 times, are among the most remarkable \HI\ features. 
The brightness and width of both \HI\ arms varies with radius: most notably, 
in the south, Arm~1 splits into three branches, followed by a broad region of 
\HI\ debris towards the west before continuing on as a single feature towards 
the north. These disturbances in the outer disk of NGC~1512 are likely caused
by tidal interaction with and accretion of the dwarf companion, NGC~1510.

Individual \HI\ clouds belonging to the NGC~1512/1510 system are found out to 
projected radii of 30\arcmin\ ($\sim$83 kpc). The velocity gradient detected 
within the extended clumps agrees with that of the neighbouring spiral arms, 
suggesting that they are condensations within the outermost parts of the disk. 
Their \HI\ properties and evidence for optical and $UV$ counterparts are 
discussed in Section~3.3.

In regions of high \HI\ column density, mostly within the arms and the bridge, 
star formation is most prominent. We will analyse the relation between the 
star formation rate and the \HI\ column density in Section~4.3.

\begin{table} 
\caption{ATCA \HI\ measurements and derived properties}
\centering 
\label{tab:ngc1512} 
\begin{tabular}{lc}
\hline
                         & NGC~1512                     \\
\hline
velocity range           & 750 -- 1070\kms              \\
$v_{\rm sys}$            & 900\kms                      \\
inclination ($i$)        & $\sim$35\degr                \\ 
position angle ($PA$)    & $\sim$260\degr               \\
$v_{\rm rot}$            & $\sim$150--200\kms           \\ 
\HI\ flux density (\FHI) & 268~Jy\kms                   \\ 
\HI\ diameter            & $\sim40\arcmin \times 30\arcmin$ \\ 
\HI/opt. diameter ratio  & $\sim$4                      \\
\HI\ mass (\MHI)         & $5.7 \times 10^9$\Msun       \\ 
dynamical mass (\Mdyn)   & $\ga 3 \times 10^{11}$\Msun  \\
\MHI/\LB                 & $\sim$1                      \\
\Mdyn/\LB                & $\sim$50                     \\
\MHI/\Mdyn               & $\sim$0.02                   \\
\hline
\end{tabular}
\end{table}

\begin{figure*} 
\label{fig:pv-diagram} 
  \mbox{\psfig{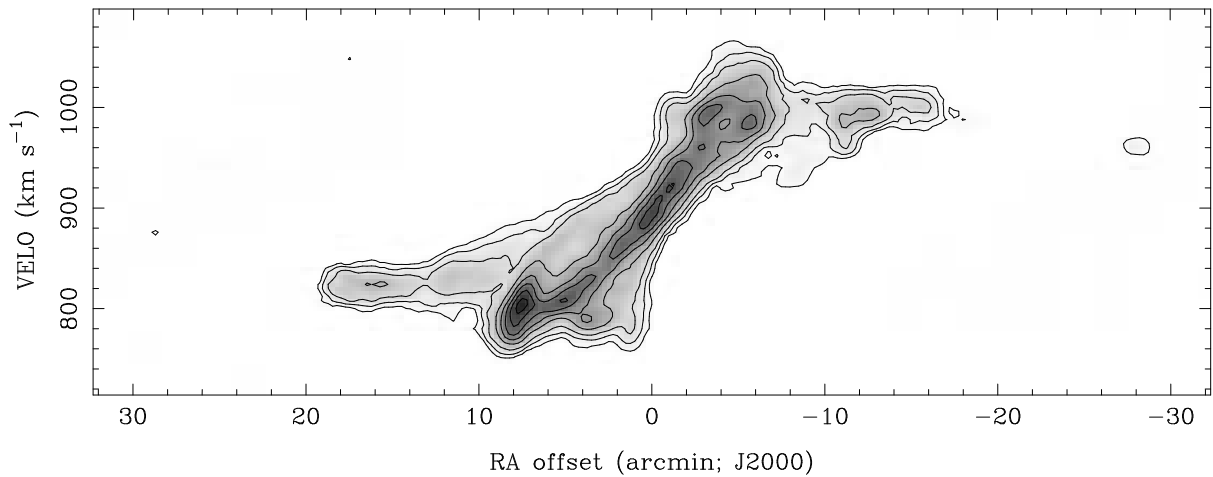}}
\caption{Major-axis \HI\ position-velocity diagram of the galaxy NGC~1512, 
   obtained from the ATCA \HI\ data shown in Fig.~4, using $PA$ = 90\degr\
   over a width of 20\arcmin. The TDG candidate, N1512-west, is just visible
   to the right.} 
\end{figure*}

\subsection{\HI\ in NGC~1512} 

The \HI\ emission from the galaxy NGC~1512 covers a velocity range from about 
750 to 1070\kms. We measure an integrated \HI\ flux density of \FHI\ = 268 
Jy\kms, which agrees very well with the HIPASS \FHI\ reported by Koribalski et 
al. (2004; see Table~1). This agreement indicates that very little diffuse \HI\
emission has been filtered out by the interferometric observation. Adopting a 
distance of 9.5 Mpc, the \HI\ flux density corresponds to an \HI\ mass of $5.7 
\times 10^9$\Msun. The majority of the detected neutral gas clearly belongs to 
NGC~1512; this is evident from the center and symmetry of the gas distribution 
and the gas kinematics. The \HI\ extent of NGC~1512 is at least a factor four
larger than its optical $B_{25}$ size. The \HI\ mass to blue luminosity ratio 
is $\sim$1. See Table~\ref{tab:ngc1512} for a summary of the galaxy properties 
as determined from the ATCA \HI\ data.

\begin{figure*} 
\label{fig:vrot} 
  \mbox{\psfig{file=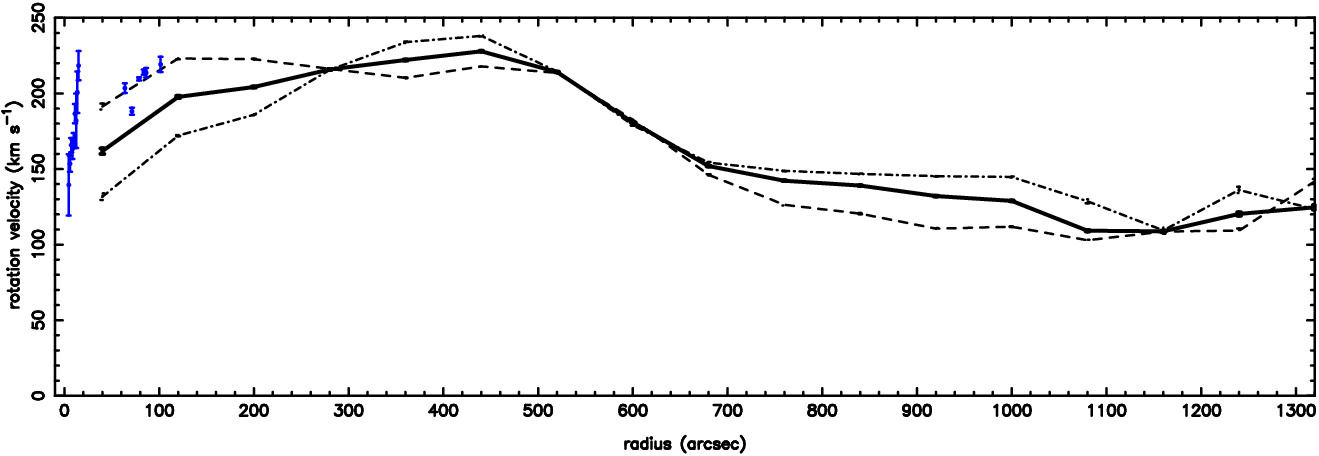,height=17cm,angle=-90}}
\caption{Rotation curve, \vrot(r), of the galaxy NGC~1512 as obtained from the 
   ATCA \HI\ velocity field shown in Fig.~2. Assuming \vsys\ = 900\kms, $PA$ 
   = 262\degr\ and an increasing inclination angle, $i$ = 30\degr\ -- 50\degr,
   we fit the approaching side (dashed line), the receding side (dash-dotted 
   line) and both sides (solid line). In the inner part ($<$ 100\arcsec) data 
   points -- with error bars -- from the \Ha\ rotation curve obtained by Buta 
   (1988) have been overlaid.} 
\end{figure*}

The interferometric \HI\ data allow us to determine the gas dynamics of the 
system. In Fig.~4 we display the \HI\ channel maps (smoothed to 20\kms\ 
resolution) which show a relatively regular rotating inner disk of NGC~1512 
and a more disturbed outer disk. Near the systemic velocity the change in 
the kinematics at a radius of $\sim$5\arcmin\ appears particularly abrupt. 
The extent and kinematics of the (spiral) arm curving towards the east are 
spectacular, nearly matched by a broader, less well-defined arm curving 
towards the west.
Another view of the galaxy kinematics is presented in Fig.~5 in form of a
major axis position-velocity ($pv$) diagram. This was obtained by summing the 
central 20\arcmin\ along $PA$ = 90\degr\ using a 3$\sigma$ cutoff to avoid 
adding excessive noise to the \HI\ signal. It shows the line-of-sight rotation 
velocity of NGC~1512 as a function of radius. The observed decrease of the
\HI\ velocities beyond a radius of 8--10\arcmin, compared to the inner disk, 
is most likely caused by an increase in the inclination of the \HI\ disk.
This warping of the outer spiral/tidal arms -- which is quite common in spiral 
galaxies -- could be related to or potentially caused by the interaction with 
NGC~1510.

\begin{figure*}
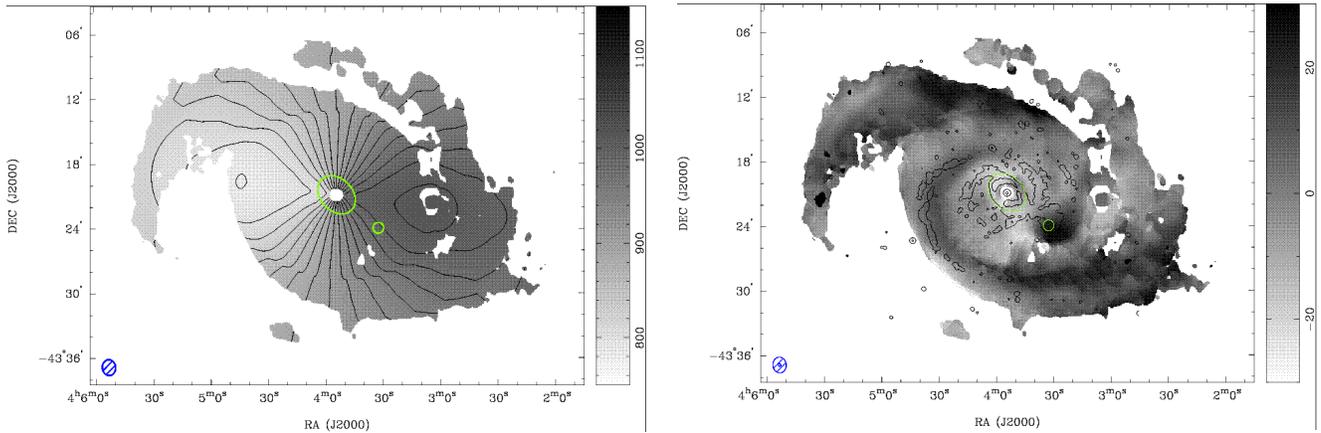
 
\label{fig:vel2fit} 
\begin{tabular}{cc}
 \mbox{\psfig{file=vel2model74.mask2.cps,height=8.5cm,angle=-90}} &
 \mbox{\psfig{file=vel2resid74+galex.fuv.cps,height=8.5cm,angle=-90}} \\
\end{tabular}
\caption{{\bf (Left)} Model velocity field resulting from the rotation curve
      fit shown in Figure~6, masked with the outer contour of NGC~1512's \HI\ 
      distribution. The contour levels range from 785 to 1025\kms\ in steps
      of 15\kms. {\bf (Right)} Residual velocity field of the interacting 
      galaxy pair NGC~1512/1510. As in Figs.~2--4, the ellipse (center) and 
      the small circle ($\sim5\arcmin$ towards the SW) mark the position and 
      size of the 20-cm radio continuum emission from NGC~1512 and NGC~1510, 
      respectively. In addition, we overlaid the GALEX FUV contours onto the
      residual \HI\ velocity field displayed on the right side.}
\end{figure*}

We used the {\sc gipsy} program {\sc rotcur} (Begeman 1987) to fit the \HI\
rotation curve of the galaxy NGC~1512. As a first step, we tried to obtain 
its centre position and systemic velocity, \vsys, using five rings within the 
inner velocity field ($r < 7\arcmin$). As the results did not converge, we 
proceeded with the GALEX peak position for NGC~1512, $\alpha,\delta$(J2000) 
= $04^{\rm h}\,03^{\rm m}$\,54\fs2, --43\degr\,20\arcmin\,56\farcs5. With this
centre position held fixed we find \vsys\ = $900 \pm 5$\kms.
We might expect the kinematic centre position of the NGC~1512/1510 system to
shift with radius from the core of NGC~1512 towards its interaction partner,
NGC~1510 (\vopt\ = 989\kms), now located $\sim$5\arcmin\ to the southwest. 
Furthermore, we find the position angle, $PA$, of NGC~1512 appears reasonably 
constant around $262\degr \pm 1\degr$, while the inclination angle, $i$, 
varies significantly. With the centre position, \vsys, and $PA$ set to the 
values given above, we find the inclination angle to increase from 
$\sim$30\degr\ ($r < 8\arcmin$) to 46\degr\ ($r > 10\arcmin$). The latter 
is consistent with the apparent change in the ellipticity (i.e. increasing 
major to minor axis ratio) of NGC~1512's gas distribution with radius.

The resulting rotation curve, $v_{\rm rot}(r)$, is shown in Fig.~6. The 
maximum rotational velocities of \vrot\ $\approx$ 225\kms\ are reached at 
radii between $\sim$300\arcsec\ and $\sim$500\arcsec. Beyond that \vrot\ 
rapidly decreases, reaching $\sim$110\kms\ at $r$ = 1200\arcsec\ (55 kpc). 
The residual velocity field (see Fig.~7) shows deviations up to approximately
$\pm$30\kms, most notably near the position of NGC~1510 and in the outer
spiral/tidal arms. The inner disk also shows deviations along an eastern arc 
(similar to a one-armed spiral) which roughly agrees with the elongated 
star-forming spiral arm of NGC~1512 seen in the GALEX $UV$ images. The 
passage of the 
companion would have unsettled the mass distribution, possibly causing a
density wave or one-armed spiral (as seen in the residual velocity field).

We estimate a dynamical mass of about $3 \times 10^{11}$\Msun\ for NGC~1512,
based on a galaxy radius of $r$ = 55 kpc and a rotational velocity of \vrot\ 
= 150\kms. If the outer \HI\ clouds at $r$ = 83 kpc are bound to NGC~1512, 
the dynamical mass increases to $4.3 \times 10^{11}$\Msun.

We note that the rotation velocity of NGC~1512 is comparable or higher (200
-- 220\kms) in the nuclear ring than in the inner disk, and significantly 
higher than in the outer \HI\ envelope. The \MHI\ to \Mdyn\ ratio indicates 
that $\la$2\% of the mass of NGC~1512 is in the form of \HI\ gas. No estimate 
of the molecular gas mass in NGC~1512 or NGC~1510 is currently available. 

\begin{figure*} 
\label{fig:tdg} 
\begin{tabular}{ccc}
  \mbox{\psfig{file=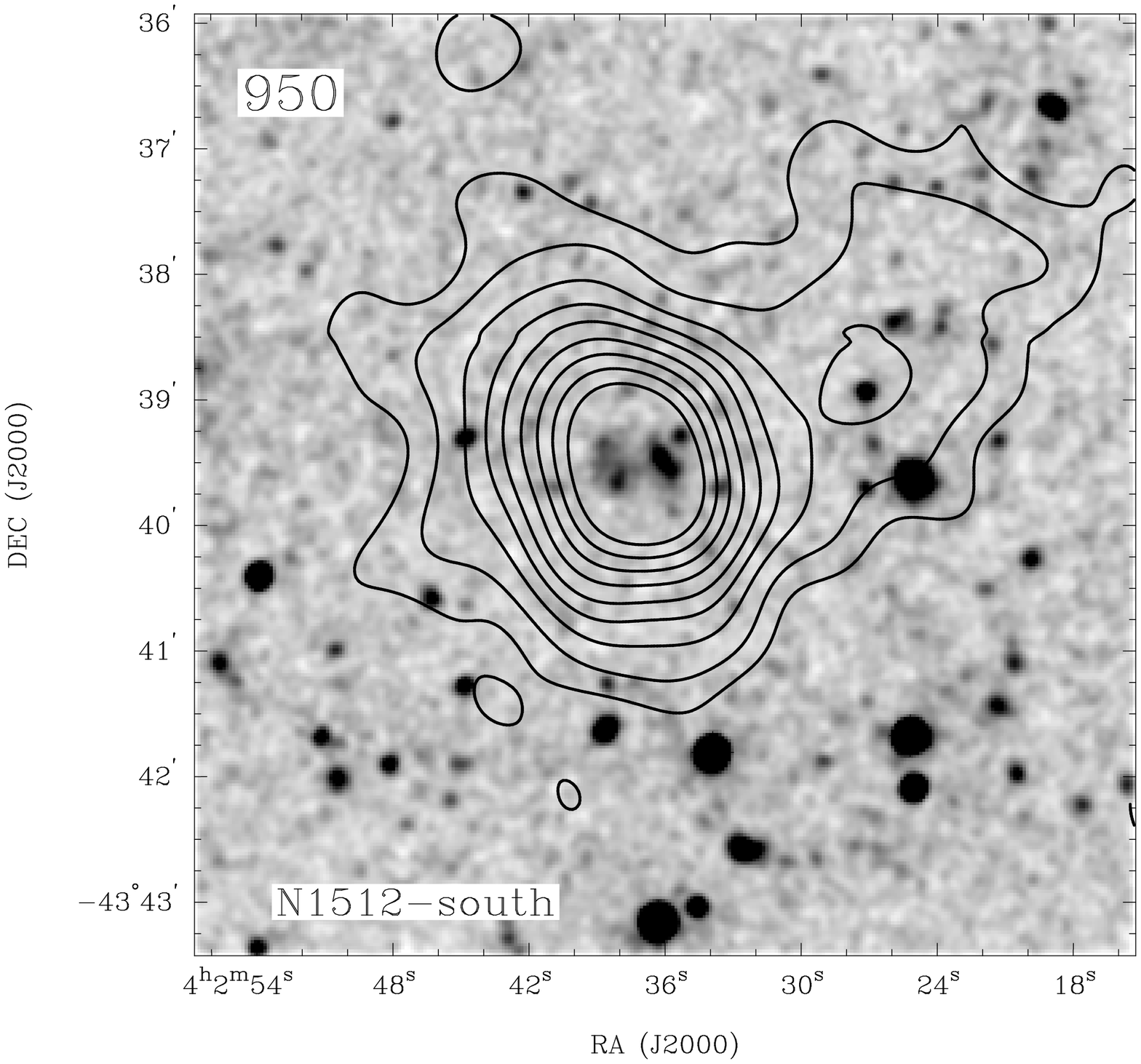,width=5cm,angle=-90}} &
  \mbox{\psfig{file=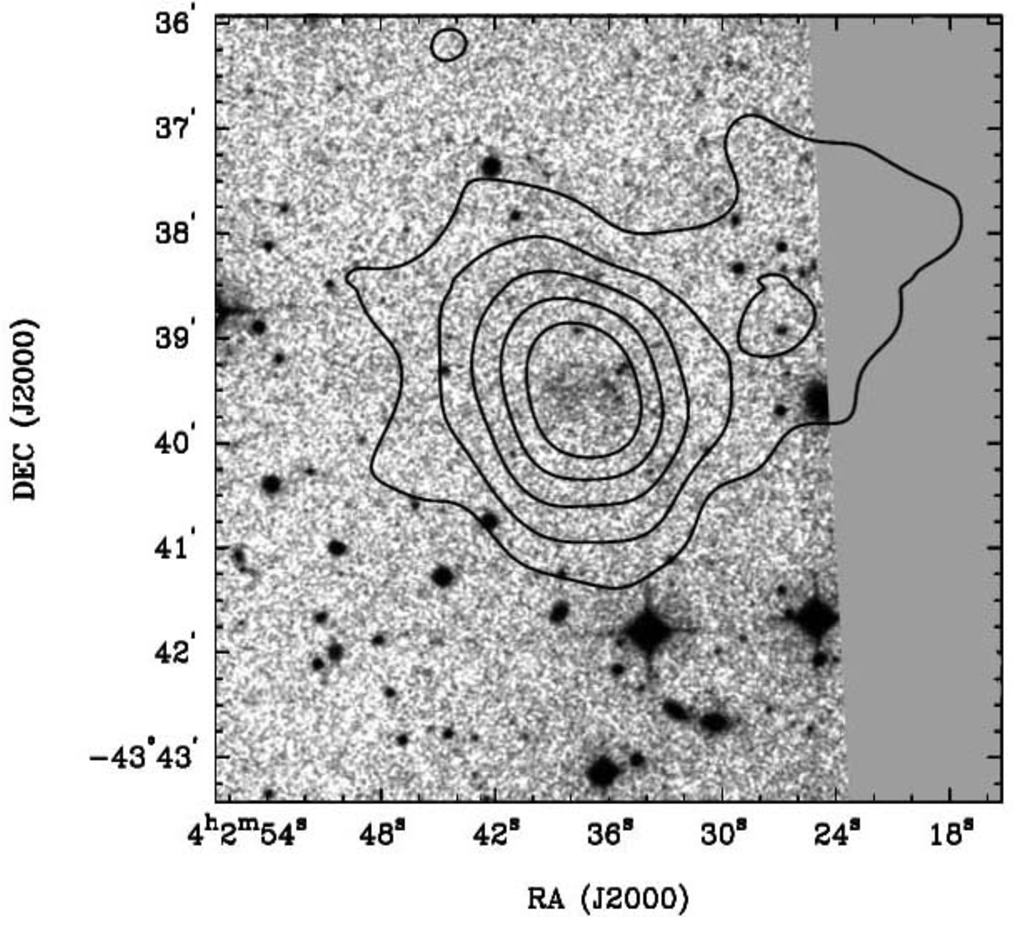,width=5cm,angle=-90}} & 
  \mbox{\psfig{file=N1512-south.mbspect.ps,width=4.5cm,angle=-90}} \\
  \mbox{\psfig{file=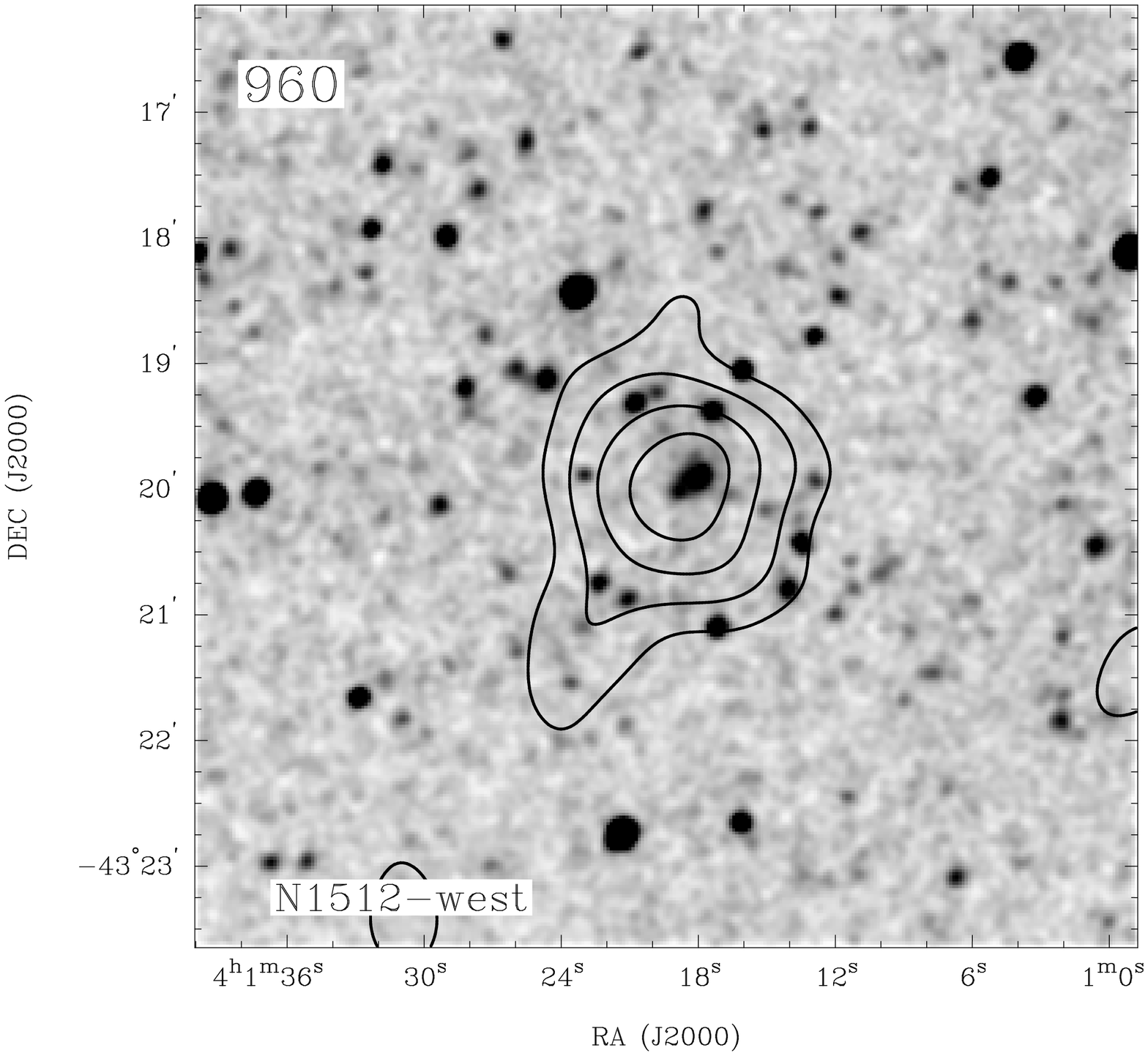,width=5cm,angle=-90}} &
  \mbox{\psfig{file=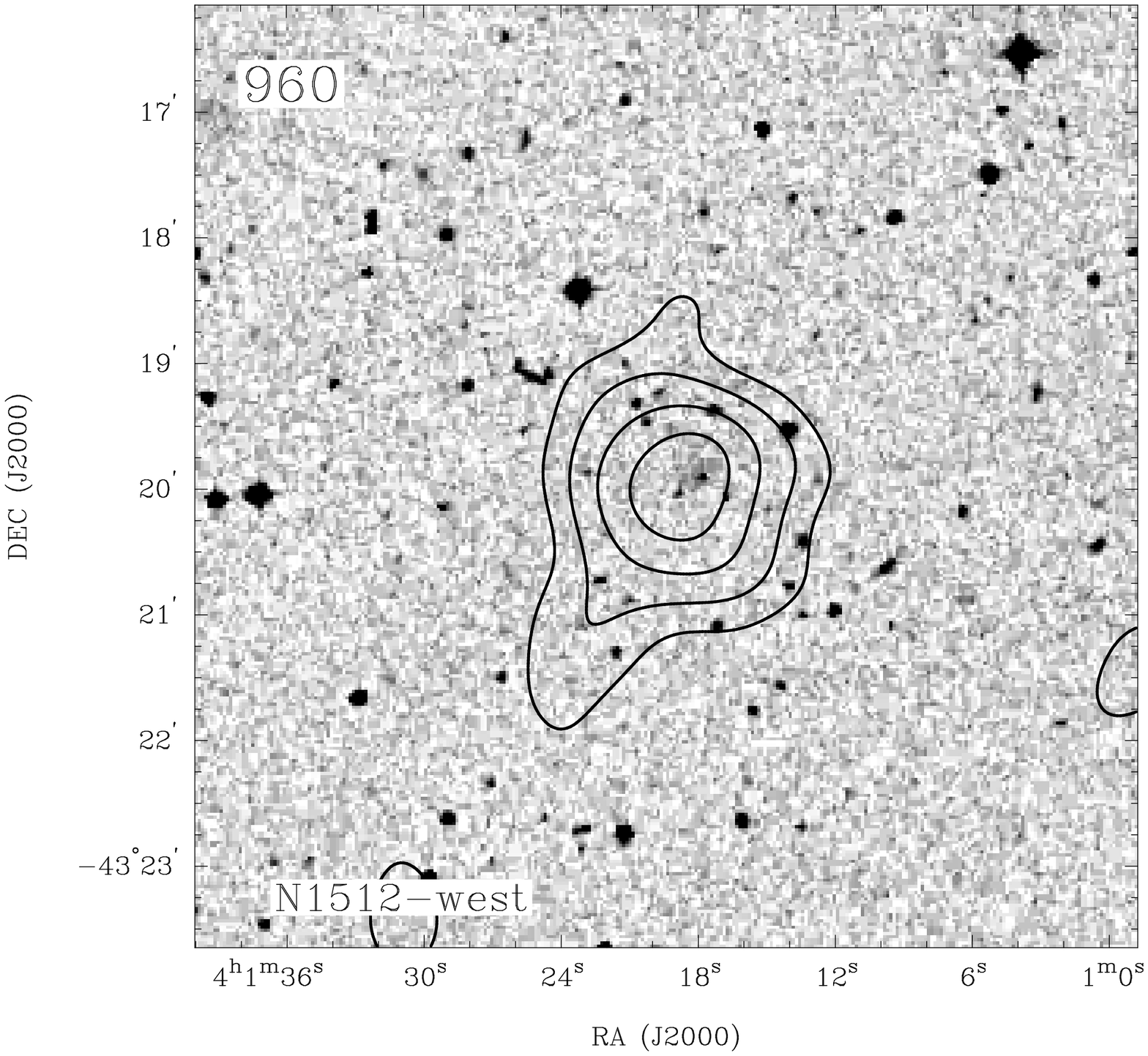,width=5cm,angle=-90}} & 
  \mbox{\psfig{file=N1512-west.mbspect.ps,width=4.5cm,angle=-90}} \\
\end{tabular}
\caption{Multi-wavelength images and ATCA \HI\ spectra of the two {\em tidal 
   dwarf galaxy} candidates in the NGC~1512/1510 system; for details see 
   Section~3.3. Black contours (--5, 5, 8, 12, 16, 20, 25, 30, 35 and 40
   mJy\,beam$^{-1}$) show the ATCA \HI\ emission at 950\kms\ for N1512-south 
   (top) and at 960\kms\ for N1512-west (bottom), while the greyscale depicts 
   the stellar populations as seen in the GALEX $NUV$ image {\bf (left)} and 
   the respective optical images {\bf (right)}. Malin's deep optical image is 
   used for N1512-south and a DSS2 image for N1512-west.}
\end{figure*}

\begin{figure*} 
\label{fig:radio1} 
  \mbox{\psfig{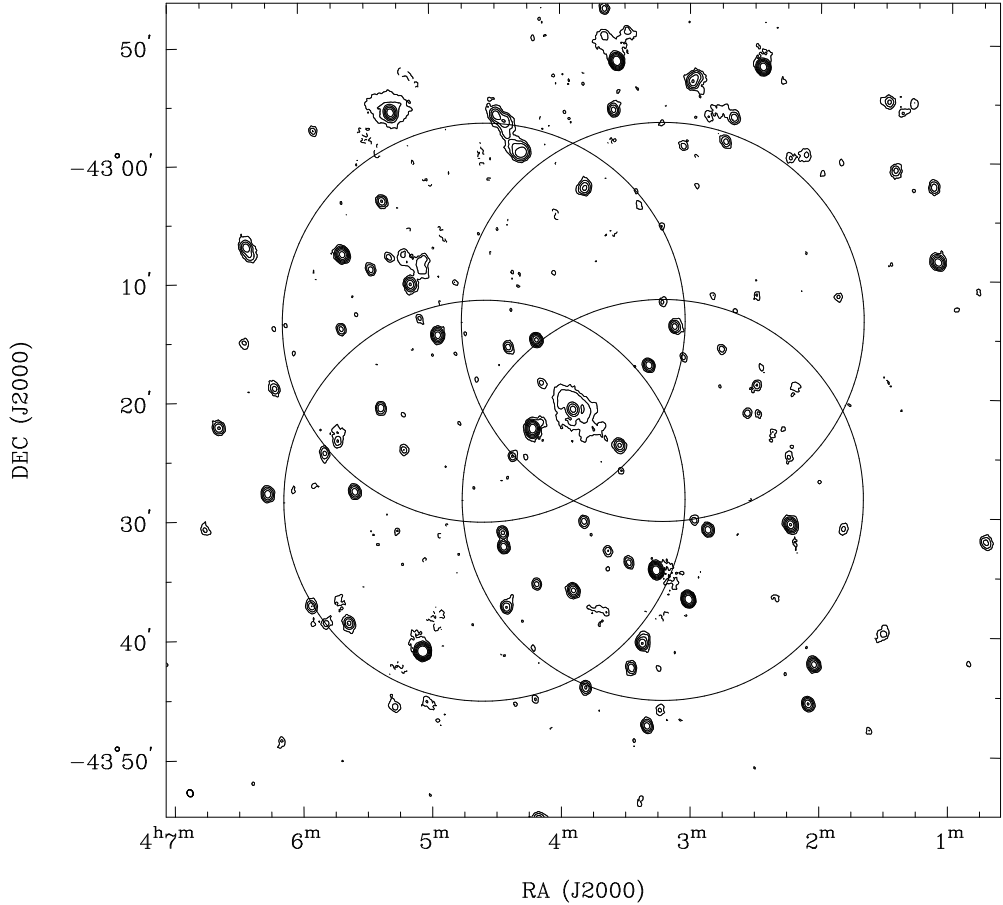}} 
\caption{20-cm radio continuum emission of the galaxy pair NGC~1512/1510 and 
   its surroundings as obtained from the ATCA wide-band data using `robust' 
   weighting. The contour levels are --0.3, 0.3, 0.6, 1.2, 2.4, 4.8, and 9.6 
   mJy\,beam$^{-1}$. The synthesised beam ($36\farcs6 \times 30\farcs0$) is 
   displayed in the bottom left corner. --- The overlaid circles (diameter = 
   33\farcm7) indicate the primary beam FWHM at the four pointing centres.}
\end{figure*}

\subsection{\HI\ in NGC~1510} 

NGC~1510 lies at a projected distance of $\sim$5\arcmin\ (13.8 kpc) from the 
centre of NGC~1512. This places it well inside NGC~1512's \HI\ disk, which 
shows an enhancement of the \HI\ column density at the position of NGC~1510. 
The offset in the residual velocity field of NGC~1512 (see Fig.~7) also 
suggests that NGC~1510 contains a small amount of \HI\ gas and/or left the 
signature of its interaction with the inner disk of NGC~1512. Gallagher et 
al. (2005) speculate that NGC~1510 may have captured gas from NGC~1512,
contributing to its enhanced SF activity. Assuming the \HI\ distribution of 
NGC~1510 is unresolved in the \HI\ maps shown here, we measure an \HI\ flux 
density of \FHI\ $\approx$ 2 Jy\kms, corresponding to an \HI\ mass of \MHI\ 
$\approx 4 \times 10^7$\Msun. This estimate is very uncertain and does not 
account for any gas NGC~1510 may have lost during the interaction, and is 
likely to include some \HI\ from the disk of NGC~1512. The estimated \HI\ 
mass (if correct) is less than 1\% of NGC~1512's \HI\ mass. The \HI\ mass to 
blue luminosity ratio of NGC~1510 would be $\sim$0.07, within the range (0.02 
-- 0.8) observed for BCD galaxies (the average ratio is $\sim$0.3; Huchtmeier 
et al. 2005, 2007).

At the position of NGC~1510, the \HI\ emission ranges from $\sim$970 to 
1060\kms. Note that the optical systemic velocity of NGC~1510 is $990 \pm 
23$\kms\ (de Vaucouleurs et al. 1991, Lindblad \& J\"ors\"ater 1981), about 
100\kms\ higher than that of NGC~1512.

\subsection{Tidal Dwarf Galaxy Candidates} 

Individual \HI\ clouds belonging to the NGC~1512/1510 system are found out to 
projected radii of $\sim$30\arcmin\ (83 kpc). Their velocities agree with the 
general rotation of the disk gas (see Figs.~2 \& 4), suggesting that these 
clumps are condensations within the outermost parts of the disk. It is likely 
that they were or are embedded in a very low surface brightness disk that 
remains undetected in our observations.

In the following we study the three most isolated \HI\ clouds which lie 
roughly along an extension of the eastern-most \HI\ arm. The first cloud,
at $\alpha,\delta$(J2000) = $04^{\rm h}\,04^{\rm m}$\,20\fs4, 
--43\degr\,34\arcmin\,16\farcs7 ($\sim$875\kms), is located only 14\farcm2 
(39 kpc) from the center of NGC~1512. We measure an approximate \HI\ flux 
density of 0.56 Jy\kms. No stellar counterpart is detected.

The second cloud, which is the brightest and most extended of the three \HI\ 
cloud, is located at $\alpha,\delta$(J2000) = 
$04^{\rm h}\,02^{\rm m}\,37^{\rm s}$, --43\degr\,39\arcmin\,32\arcsec\ (\HI\ 
peak position), 23\farcm3 (64 kpc) from the center of NGC~1512. It has an \HI\ 
peak flux of $\sim$1 Jy\,beam$^{-1}$ and a total \HI\ flux density $\sim$3 
Jy\kms\ (\MHI\ = $6 \times 10^7$\Msun). There is a clear velocity gradient 
along the cloud (940 -- 970\kms) which agrees with the general rotation pattern
of NGC~1512; its centre velocity is $\sim$950\kms. The deep optical image (see
Fig.~8, top right) reveals the faint optical counterpart\footnote{The 
   deep optical image of the NGC~1512/1510 system, obtained by David Malin, 
   is partially shown in our Fig.~1 and is, in its full size ($\sim40\arcmin 
   \times 50\arcmin$), available at 
   www.aao.gov.au/images/deep\_html/n1510\_d.html .}:
a diffuse knot coinciding with 
the \HI\ maximum, barely visible in the second-generation Digitised Sky Survey 
(DSS2). In addition, we find clear evidence of star formation in the GALEX 
$FUV$ and $NUV$ images (see also Section~4.2). Fig.~8 (top) shows the locations
of the optical and ultraviolet emission with respect to the \HI\ emission. We 
suggest that the core of this \HI\ cloud (with an \HI\ mass of $\sim2 \times 
10^7$\kms) is a {\em tidal dwarf galaxy} (TDG) and refer to it as N1512-south.

The third \HI\ cloud is located at $\alpha,\delta$(J2000) = 
$04^{\rm h}\,01^{\rm m}\,19^{\rm s}$, --43\degr\,20\arcmin\,03\arcsec\
($\pm$10\arcsec), 28\farcm2 (78 kpc) from the center of NGC~1512. --- Note 
that this puts it near the edge of the field well-mapped by our four 
overlapping ATCA pointings, and \HI\ sensitivity is reduced; primary beam
correction is tapered to avoid excessive noise. --- We suggest that this 
compact \HI\ cloud is a second, more evolved {\em tidal dwarf galaxy} in 
the NGC~1512/1510 system and refer to it as N1512-west. It has an \HI\ flux 
density of at least $\sim$0.3 Jy\kms\ (\MHI\ $\ga 0.6 \times 10^7$\Msun) and 
a centre velocity of $\sim$960\kms. Unfortunately, Malin's deep optical image 
does not extend this far west of NGC~1512. Nevertheless, clear evidence of 
star formation is again found in the GALEX $FUV$ and $NUV$ images (see Fig.~8,
bottom). 

Both N1512-south and N1512-west are detected in the highest resolution 
($13\farcs0 \times 11\farcs3$) ATCA \HI\ cubes. We find \HI\ peak fluxes 
$\sim$10 mJy\,beam$^{-1}$ over 15--30\kms. The resulting \HI\ column 
densities are $\ga$10$^{21}$ atoms~cm$^{-2}$ (or $\ga$8\Msun\,pc$^{-2}$), 
i.e. near the local star formation threshold (e.g., Skillman 1987).
\HI\ data with higher angular \& velocity resolution (and sensitivity) are 
needed to study the TDG candidates in more detail.

After careful calibration, we estimate $FUV-NUV$ colors of $\sim$0.35--0.43 mag
for N1512-south (there are two distinct star forming knots within this TDG) 
and $\sim$0.45 mag for N1512-west (see also Section~4.2). Despite the large 
uncertainties ($\sim$0.2 mag) in the color estimates in these areas (which 
are outside the published images by Gil de Paz et al. 2007a), we find 
that the derived average ages of the detected stellar populations within the 
TDGs are no younger than 150 Myr and possibly as old as $\sim$300 Myr. 
N1512-west appears to be slightly older (more evolved) than N1512-south which 
would be expected given its compactness and location at the outermost tip of 
the extrapolated eastern arm.

Following Braine et al. (2004) we can estimate the expected molecular gas 
mass (as traced by CO emission) of TDGs to be less than 30\% of the \HI\ 
mass, i.e.  $\la2-3 \times 10^6$\Msun\ (for N1512-west). Assuming a velocity 
width of 10\kms\ (approx. half the width found in \HI), a CO peak flux of 
30 mK\,beam$^{-1}$ would be expected with a typical single dish telescope. 

If the star-forming \HI\ clumps in the outskirts of the NGC~1512/1510 are
indeed TDGs, we would expect to detect \Ha\ emission from stars recently
formed in the tidal debris. This young stellar population must exist in
addition to the (on average) older population inferred from the GALEX 
$FUV-NUV$ colors.

\begin{figure*}
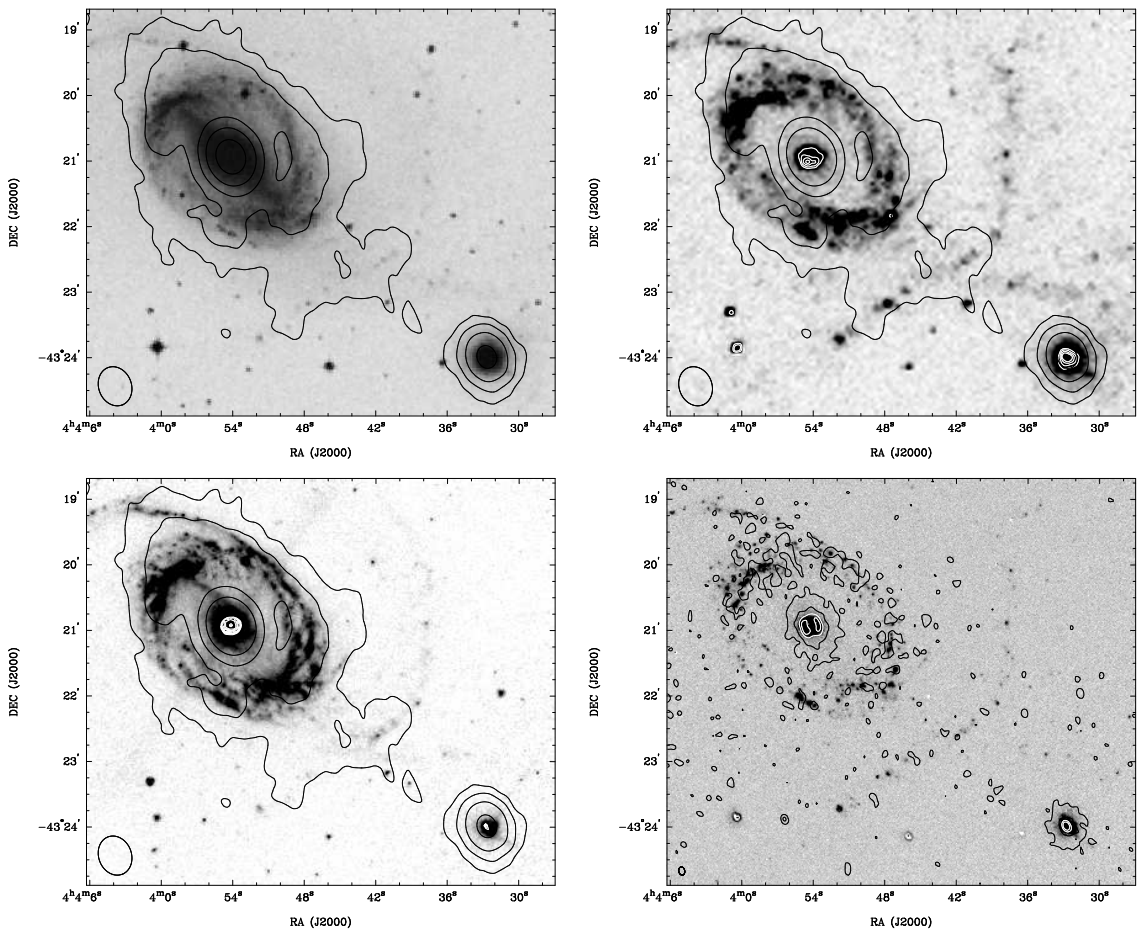
 
\label{fig:radio2} 
\begin{tabular}{cc}
  \mbox{\psfig{file=ngc1512.1384.r0.icln.regrid+dss2blue.ps,width=6cm,angle=-90}} &
  \mbox{\psfig{file=ngc1512.1384.r0.icln.regrid+nuv.ps,width=6cm,angle=-90}}  \\
  \mbox{\psfig{file=ngc1512.1384.r0.icln.regrid+8micron.ps,width=6cm,angle=-90}}  & 
  \mbox{\psfig{file=ngc1512.1384.u+6.icln.regrid+ha.ps,width=6cm,angle=-90}}  \\
\end{tabular}
\caption{Zooming-in towards the galaxy pair NGC~1512 and NGC~1510. 
   {\bf (Top left)} ATCA 20-cm radio continuum map (black contours as in 
   Fig.~9) overlaid onto the DSS2 optical $B$-band image,
   {\bf (top right)} overlaid onto the GALEX $NUV$ image, and
   {\bf (bottom left)} overlaid onto the SINGS 8$\mu$m image.
   {\bf (Bottom right)} High resolution ATCA 20-cm radio continuum map (black 
   contours) overlaid onto the SINGG \Ha\ image. The contour levels are 0.09, 
   0.25, 0.65, and 0.95 mJy\,beam$^{-1}$; the synthesised beam is $7\farcs6 
   \times 5\farcs5$. In those panels where the nuclear star-forming ring is 
   resolved, white contours have been added to reflect this. --- In the bottom 
   left corner of each panel we show the respective ATCA synthesised beams.}
\end{figure*}

\begin{figure*} 
\label{fig:multicolor}
 \mbox{\psfig{file=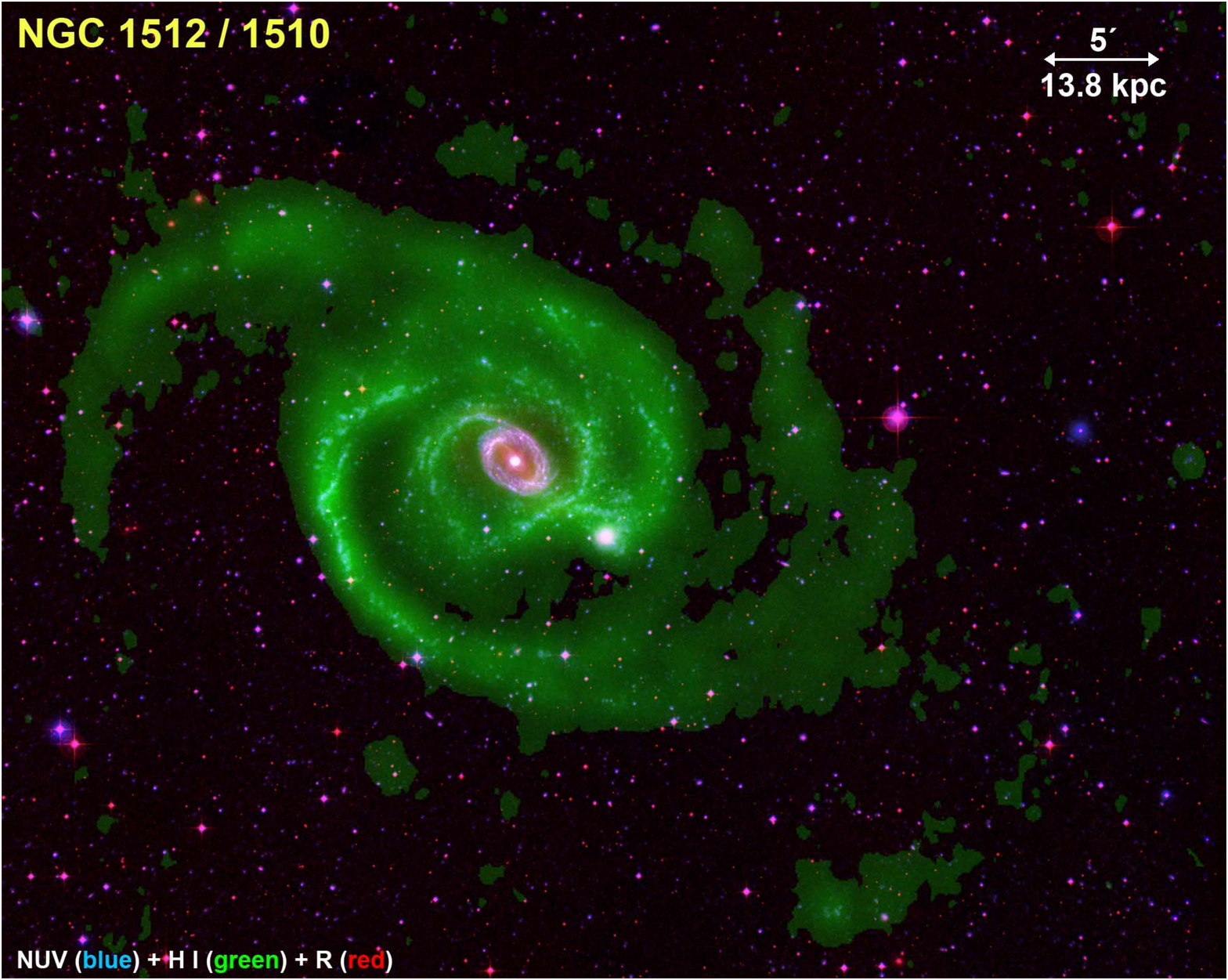,width=18cm}} 
\caption{Multi-wavelength color-composite image of the galaxy pair 
   NGC~1512/1510 obtained using the DSS $R$-band image (red), the ATCA \HI\ 
   distribution (green) and the GALEX $NUV$-band image (blue). The Spitzer 
   24$\mu$m image was overlaid just in the center of the two galaxies. We
   note that in the outer disk the $UV$ emission traces the regions of 
   highest \HI\ column density.}
\end{figure*}

\subsection{20-cm Radio Continuum Emission} 

Figure~9 shows the 20-cm radio continuum emission towards the galaxy pair 
NGC~1512/1510 and its surroundings. Both galaxies are clearly detected. The 
field contains a large number of unresolved radio sources as well as a few 
head-tail and wide-angle tail radio galaxies (incl. PMN J040150.3--425911).

The barred spiral galaxy NGC~1512 shows extended continuum emission ($\sim
5\arcmin \times 3\arcmin$) and a bright core while the much smaller BCD galaxy 
NGC~1510 appears unresolved at 30\arcsec\ resolution. The {\em `fish'} shaped 
radio continuum emission is due to enhanced star-formation in the region 
between the two galaxies which corresponds to the {\em tail} of the {\em fish}.
NGC~1512's inner star-forming ring which is well defined in the optical-, \Ha- 
and $UV$-images is embedded within the radio continuum emission (see Fig.~10). 
At high resolution, our surface brightness sensitivity is insufficient to fully
map the structure of the inner ring, but NGC~1512's nuclear ring is clearly 
detected as well as the extended disk of emission from NGC~1510.

To estimate the star formation rate (SFR) of a galaxy from our 20-cm data we
use two approaches: (1) the formation rate of recent, high mass stars ($M > 
5$\Msun) is calculated using SFR [\MMoy] = $0.03~D^2~S_{\rm 20cm}$ (Condon et 
al. 2002), where $D$ is the distance in Mpc and $S_{\rm 20cm}$ the 20-cm radio 
continuum flux density in Jy. We measure $S_{\rm 20cm}$ = 38.8~mJy for NGC~1512
and 4.5~mJy for NGC~1510, resulting in 0.11\Moy\ and 0.012\Moy, respectively. 
(2) In order to derive the formation rate of all stars ($M > 0.1$\Msun) we 
multiply by 4.76 (see Condon et al. 2002), resulting in 0.50\Moy\ (NGC~1512) 
and 0.06\Moy\ (NGC~1510).

\begin{figure*}
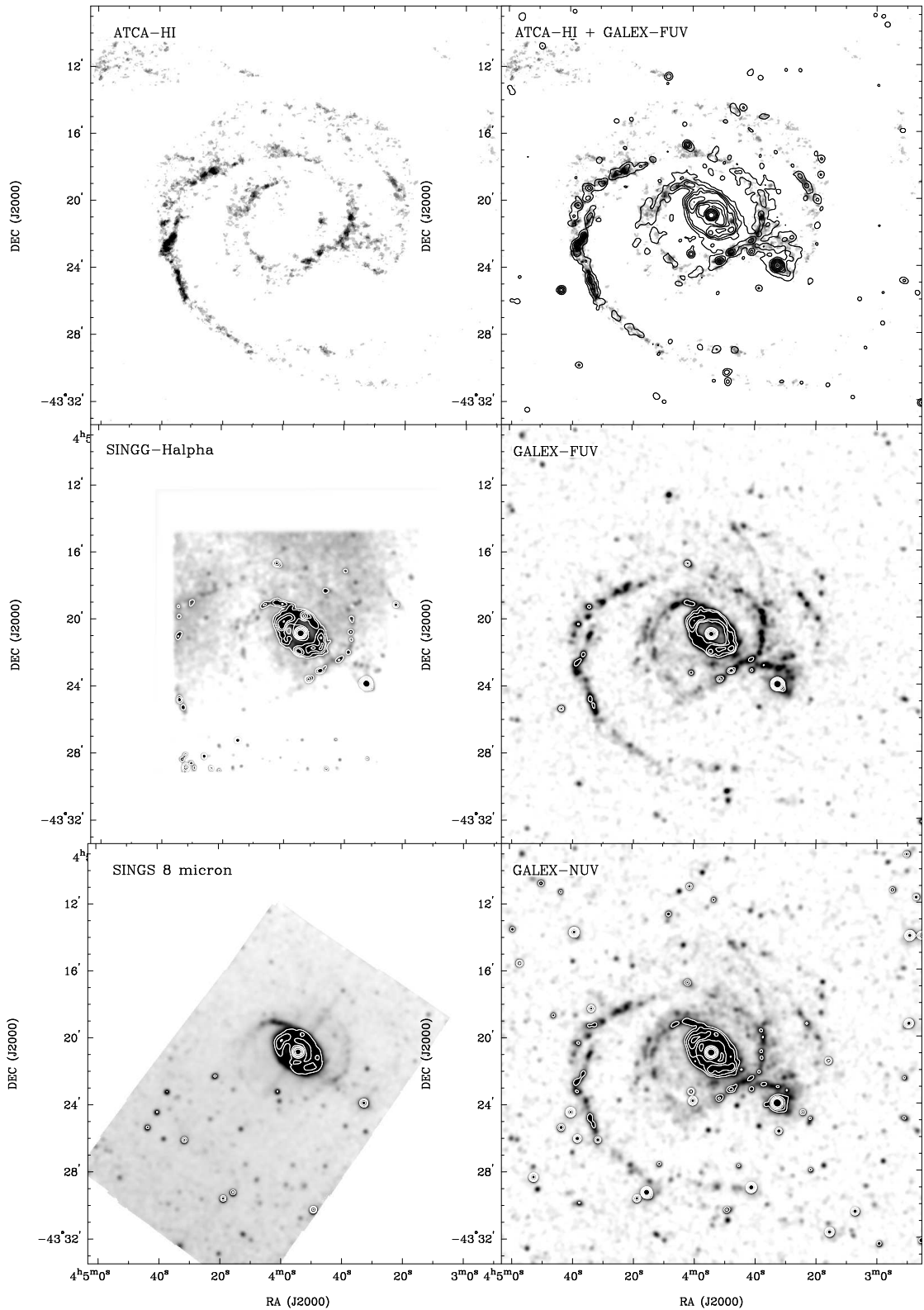
 
\label{fig:multi}
\begin{tabular}{cc}
 \vspace{-1cm}
 \mbox{\psfig{file=ngc1512.line.4km5.r0+6.15mom0.sub.ps,width=7.5cm,angle=-90}}&
 \hspace{-2cm}
 \mbox{\psfig{file=ngc1512.hi+fuv.sub.ps,width=7.5cm,angle=-90}}\\
 \vspace{-1cm}
 \mbox{\psfig{file=ngc1512.singg-ha.mask.convol15.sub.ps,width=7.5cm,angle=-90}} &
 \hspace{-2cm}
 \mbox{\psfig{file=ngc1512.fuv-large.convol15.sub.ps,width=7.5cm,angle=-90}} \\
 \mbox{\psfig{file=ngc1512_v7.phot.4.convol15.sub.ps,width=7.5cm,angle=-90}} &
 \hspace{-2cm}
 \mbox{\psfig{file=ngc1512.nuv-large.convol15.sub.ps,width=7.5cm,angle=-90}} \\
\end{tabular}
\caption{The galaxy pair NGC~1512/1510 in \HI, FUV, NUV, \Ha\ and 8$\mu$m. 
   All six panels show the same area at 15\arcsec\ (700 pc) resolution. The 
   top right panel shows the FUV emission (contours) overlaid onto the \HI\ 
   distribution (greyscale). The greyscale has been adjusted such that faint 
   features in the outer disk of the system are emphasised, while the inner 
   region is over-exposed. The white contours, in some panels, help to trace 
   the emission in the over-exposed areas.}
\end{figure*}

\section{Discussion} 

The \HI\ diameter of the galaxy NGC~1512 is large ($\sim$40\arcmin\ or 
110~kpc), but similar in size to the largest \HI\ disks found in the Local
Volume. For example, ATCA \HI\ mosaics of the spiral galaxies M\,83 
(Koribalski et al. 2009) and Circinus (Curran et al. 2008) reveal diameters 
of $\sim$70\arcmin\ (80 kpc) and $\sim$65\arcmin\ (80 kpc), respectively. 
Even larger \HI\ diameters have been found for some of the most \HI-massive 
galaxies currently known, e.g. Malin\,1 (Pickering et al. 1997), NGC~6872 
(Horellou \& Koribalski 2007), and HIZOA J0836--43 (Donley et al. 2006;
Cluver et al. 2008).
Possibly the deepest single-dish \HI\ map recently obtained with the Arecibo 
multibeam system for the nearby spiral galaxy NGC~2903 also shows a very large 
\HI\ envelope and a neighbouring \HI\ cloud (Irwin et al. 2009).

A multi-wavelength study of the grand-design spiral galaxy M\,83 (HIPASS 
J1337--29) and its neighbours --- similar to the one presented here --- is 
under way; first Parkes and ATCA \HI\ results were presented by Koribalski 
(2005, 2007) and are shown on the LVHIS webpages. 
The \HI\ gas dynamics of the Circinus galaxy (HIPASS J1413--65) were studied 
by Jones et al. (1999), using high resolution data, and more recently by 
Curran et al. (2008), using a low-resolution mosaic. Circinus appears to be 
rather isolated and is difficult to study in the optical due to its location 
behind the Galactic Plane.

\subsection{The 2X-\HI\ vs X-UV disk of NGC~1512}       

A multi-wavelength color-composite image of the NGC~1512/1510 system is shown 
in Fig.~11. The combination of the large-scale \HI\ distribution with deep 
optical and $UV$ emission maps is an excellent way to highlight the locations 
of star formation within the gaseous disk. NGC~1512's \HI\ envelope is four 
times larger than its $B_{\rm 25}$ optical size (see Table~1) and about twice 
as large as the stellar extent measured from Malin's deep optical image and 
from the GALEX $UV$ images. 

Calibrated $FUV$ (1350--1750\AA) and $NUV$ (1750--2750\AA) images are provided 
by Gil de Paz et al. (2007a) as part of the \emph{GALEX Atlas of Nearby 
Galaxies}. The data were obtained on the 29th of December 2003, with an 
exposure time in both bands of 2380 seconds. The GALEX full field-of-view 
is $\sim$1\fdg2 in diameter, and the pixel size is 1\farcs5. The GALEX 
point-spread function in the central 0\fdg5 has a FWHM of $\sim$5\arcsec.

Figure~12 gives a multi-wavelength view of the NGC~1512/1510 system, shown 
with high resolution (15\arcsec\ = 700 pc) over the main star-forming disk 
(similar in size to Fig.~1). Our main 
purpose here is to emphasize how the observed extent and distribution of stars 
and gas depend on the tracer. The \HI\ distribution is by far the largest and
extends well beyond the area shown here. The GALEX $FUV$ and $NUV$ images,
shown here smoothed to an angular resolution of 15\arcsec, trace star forming 
regions out to a radius of $\sim$10\arcmin. Malin's deep optical image (see
Fig.~1) shows a very similar distribution. We expect that a deep \Ha\ mosaic 
would also match this, as hinted at by the faint chains of \HII\ regions seen 
in the rather limited SINGG \Ha\ image.
The Spitzer 8$\mu$m image allows us to see the inner spiral arms as they 
connect to the bar, but detects no emission in the outer disk. Most obviously
missing is a map of the molecular gas in the system (e.g., as traced by 
CO(1--0) emission) which is expected to be similar to the \Ha\ image.

The large gas reservoir provides copious fuel for star formation, which should
be most prominent in areas of high column density (see Section~4.4). Given a 
high-sensitivity, high-resolution \HI\ distribution, we can pinpoint the 
locations of star forming activity in the outer disks of galaxies. 

The correspondence between regions of high \HI\ column density and bright $UV$ 
emission (see Fig.~12) is excellent throughout the extended disk of NGC~1512, 
apart from the central area which shows an \HI\ depression (but must be rich 
in molecular gas). The large majority of the observed $UV$-complexes lie in 
regions where the \HI\ column density is above $2 \times 10^{21}$ 
atoms~cm$^{-2}$ (as measured in the high-resolution \HI\ map). For comparison,
the dwarf irregular galaxy ESO215-G?009 has a very extended \HI\ disk (Warren
et al. 2004) but no signs of significant star formation in the outer disk; 
its \HI\ column density reaches above $10^{21}$ atoms~cm$^{-2}$ in only a 
few locations.

Deep GALEX images of nearby galaxies show that the $UV$ profiles of many spiral
galaxies extend beyond their \Ha- or $B_{\rm 25}$ optical radius (Thilker et 
al. 2005; Gil de Paz et al. 2005, 2007b). In fact, Zaritsky \& Christlein 
(2007) suggest that $XUV$-disks exist in $\sim$30\% of the local spiral galaxy 
population. We contend that these spectacular $XUV$-disks must be located 
within even larger \HI\ envelopes, here called $2X$-\HI\ disks, which provide
the fuel for continued star formation. Ultimately, it may just be a question 
of sensitivity that limits our observations of the outer edges of stellar and
gaseous disks. We note that Irwin et al. (2009) detect \HI\ gas out to column 
densities of $3 \times 10^{17}$ atoms~cm$^{-2}$ (assuming the gas fills the 
270\arcsec\ beam).

\begin{figure*} 
\label{fig:uvcolor1} 
  \mbox{\psfig{file=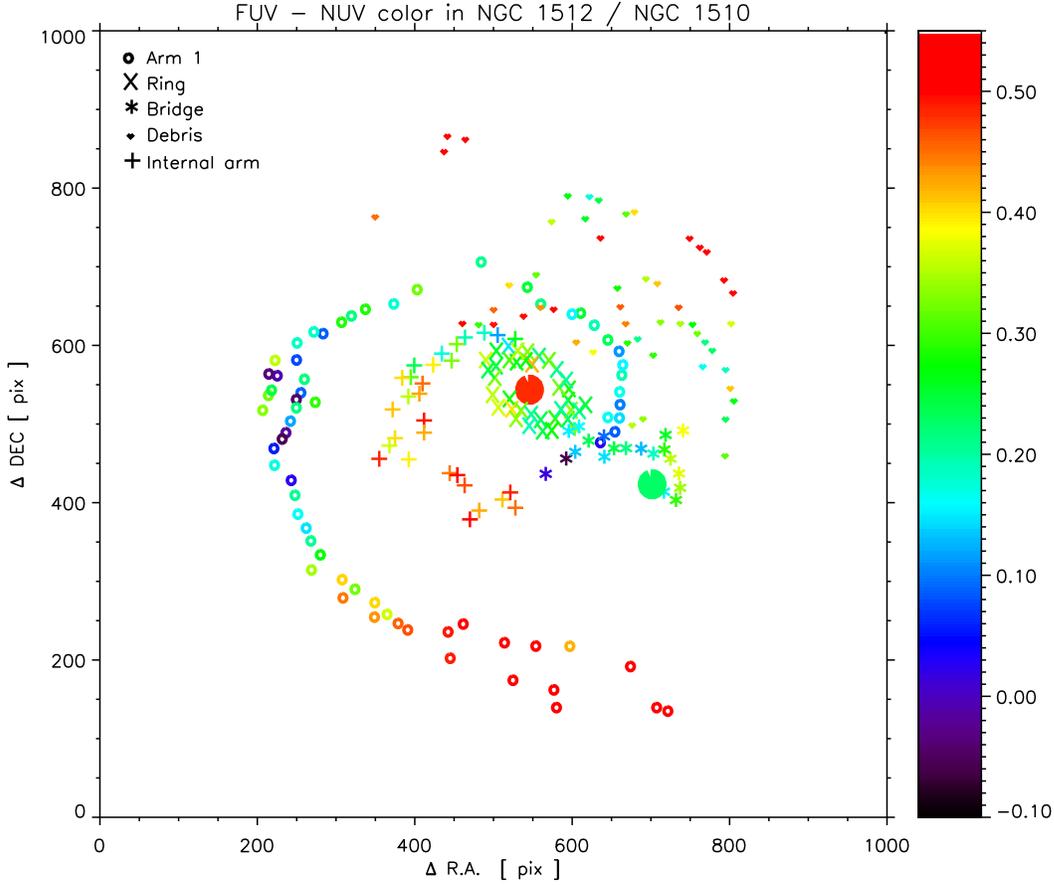,width=12cm,angle=-270}} 
\caption{GALEX $FUV-NUV$ color of the brightest $UV$-emission regions detected 
   in the NGC~1512/1510 system. The cores of both galaxies are shown as two 
   big filled circles. Different symbols denote star clusters belonging to 
   distinct regions within the system.}
\end{figure*}

\subsection{Stellar cluster ages} 

We use the GALEX $FUV$ and $NUV$ images to estimate the ages of the $UV$-rich 
star clusters in the NGC~1512/1510 system. This is done by integrating the 
counts per second (CPS) in $\sim$200 selected regions ($>$100 arcsec$^2$, 
average size = 320 arcsec$^2$) using the same polygon for both images and 
applying $m_{\lambda} = -2.5\log (CPS) + a_{\lambda}$ (Morrissey et al. 2005), 
where $a_{FUV}$ = 18.82 mag and $a_{NUV}$ = 20.08 mag (all magnitudes are 
expressed in the AB system). We did not correct for extinction which is 
negligible when computing the $FUV-NUV$ color: $A_{\rm FUV}-A_{\rm NUV} = 
-0.1~E(B-V) = -0.0011$.  

Figure~13 shows the spatial distribution and color of the analysed star 
clusters. We use different symbols to identify 
five distinct areas within the system: the ring, the internal arm, the 
bridge to NGC~1510, the western debris and Arm~1 (see Fig.~1). The $FUV-NUV$ 
colors (blue to red) range from --0.06 (youngest stellar population) at the 
southern end of the bridge to +0.68 (oldest stellar population) for the 
farthest cluster in the NW region. Uncertainties in the color estimates
strongly depend on the brightness of the star clusters ($\sim$0.06 for the 
brightest and $\sim$0.50 for the weakest objects). For the analysed clusters 
in the NGC~1512/1510 system we adopt an uncertainty of $\pm$0.20.

As extinction is negligible when computing the $FUV-NUV$ colors (see above),
higher values correspond to older ages for the last star-forming burst hosted 
by the $UV$-rich clusters. We have used the same procedure as described in 
Bianchi et al. (2005) and Hibbard et al. (2005) to estimate the age of the 
last star-forming event, assuming an instantaneous burst, and evolutionary 
synthesis models provided by Bruzual \& Charlot (2003). Table~\ref{tab:galex} 
lists the results obtained for distinct areas.

\begin{table} 
\caption{GALEX $UV$ properties of the star clusters in distinct regions 
   within the NGC~1512/1510 system.}
\label{tab:galex} 
\begin{tabular}{lrccc}
\hline
                 & \multicolumn{3}{c}{GALEX $FUV-NUV$}   \\
Region           &  min.  & max. &  median       &  Age (Myr)    \\
\hline
NGC~1512 core    &        &      &     0.48      &    $\sim$320  \\
NGC~1510         &        &      &     0.24      &    $\sim$150  \\
Ring             &   0.15 & 0.42 & $0.28\pm0.06$ & 130--180--300 \\
Bridge           & --0.06 & 0.39 & $0.21\pm0.12$ &  10--120--270 \\
Int. Arm         &   0.12 & 0.51 & $0.38\pm0.11$ &  80--270--330 \\
Arm~1            & --0.05 & 0.66 & $0.27\pm0.19$ &  10--170--380 \\
Debris           &   0.17 & 0.68 & $0.39\pm0.13$ & 100--270--390 \\
\hline
\end{tabular}
\end{table}

While the $UV$ colors suggest that the average stellar population in the 
core of NGC~1512 (red circle) is about twice as old as that of NGC~1510 (green
circle), the high \Ha\ emission in both galaxies also indicates significant
recent star formation. We conclude that NGC~1512 and NGC~1510 contain both 
a young stellar population and an older, more evolved stellar population.

As shown in Fig.~13, there are definite color gradients along the spiral arms 
and other regions within the NGC~1512/1510 system. For example, while regions 
within the inner star-forming ring of NGC~1512 generally have similar colors, 
$(FUV-NUV)_{\rm ring} = 0.28 \pm 0.06$ (age $\sim$180~Myr), slightly younger
ages are found towards both ends of the bar, i.e. at the start of the inner 
arms. Regions located within the bridge between NGC~1512 and NGC~1510 are -- 
on average -- even younger, $(FUV-NUV)_{\rm bridge} = 0.21 \pm 0.12$ (age 
$\sim$120~Myr), with ages of $\sim$10~Myr (the youngest regions in the whole 
system) near NGC~1512, and $\sim$270~Myr near NGC~1510. As shown in Fig.~12, 
$UV$-bright regions close to NGC~1512 coincide well with the \HI\ column 
density maxima; their young derived age is consistent with the \Ha\ emission 
found in these knots. A $UV$-color gradient is also observed along the 
prominent eastern arm (Arm~1). In the eastern-most regions, which also show 
some \Ha\ emission, we measure $FUV-NUV$ colors around --0.05 (age $\sim$10
Myr). As the arm curves towards the south, the $UV$-rich clusters appear to 
get older, reaching $FUV-NUV \sim 0.66$ (age $\sim$380~Myr) within the two 
streams of the outermost regions. 

Within the debris in the NW area, stellar clusters near NGC~1510 tend to have 
younger ages ($FUV-NUV \sim 0.2-0.3$, ages of 100--200~Myr) than those located 
towards the north of NGC~1512 ($FUV-NUV \sim 0.5-0.7$, ages of 300--400~Myr).
The broadening of the \HI\ spiral arm at the position of the NW debris 
suggests that something has dispersed both the neutral gas and the stellar 
component in this region. The age gradient found in the stellar clusters 
indicates that it probably is due to the gravitational interaction with the 
BCD galaxy NGC~1510. 

The overall gas distribution together with the star formation history of the 
system provides some hints as to the gravitational interaction between the 
large spiral NGC~1512 and the BCD galaxy NGC~1510 and its effects on the 
surrounding medium. The youngest star forming regions are found mostly to the 
east \& west of NGC~1512, while regions towards the north \& south of NGC~1512 
are generally older. This east-west (young) versus north-south (old) symmetry 
might indicate the passage of NGC~1510 as it is accreted by NGC~1512. This 
interaction might have (1) triggered the bar in NGC~1512 (unless this was the 
result of previous interactions or minor mergers) causing gas within the 
co-rotation radius to flow towards the nuclear region, thus providing fuel
for continuous star formation, (2) affected the spiral arm pattern causing
broadening and splitting as well as enhanced star formation, and (3) led to
the ejection of material to large radii where it may become unbound, forming
dense clumps able to form new stars. Evidence of the latter is the observation
of two {\em tidal dwarf galaxies} at the outermost regions of the NGC~1512/1510
system.

\begin{figure} 
\label{fig:sfr}
 \mbox{\psfig{file=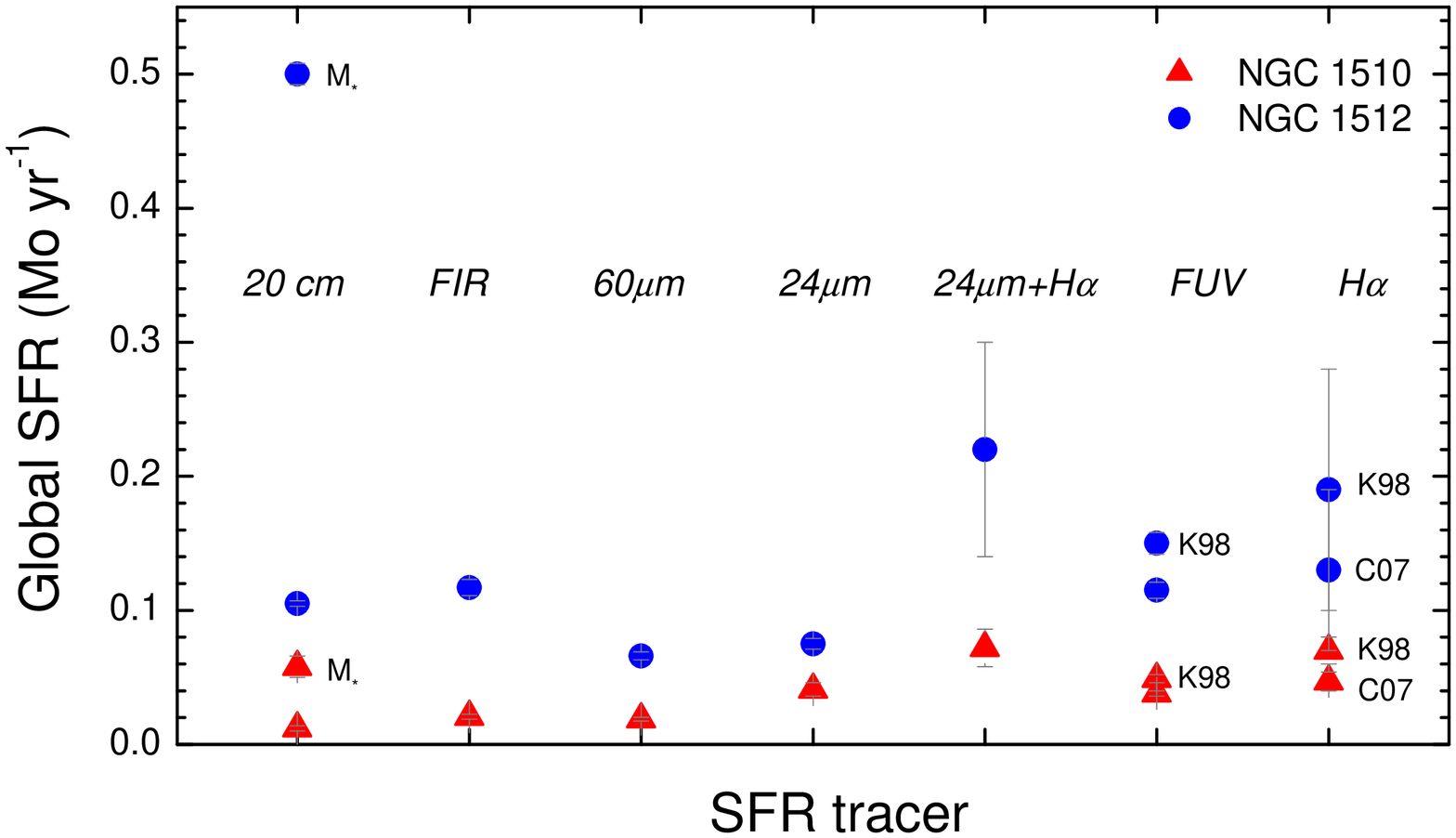,width=4.5cm,angle=-90}} 
\caption{Global star formation rates (SFR) of the galaxies NGC~1512 (blue 
   circles) and NGC~1510 (red triangles) as derived from measurements at 
   various wavelengths. All SFR estimates are listed in Table~5 and described 
   in the text. The SFR tracers are arranged along the x-axis roughly by the 
   average age of the contributing star burst population ($\sim$10~Myr for 
   \Ha; $\sim$100~Myr for 20-cm radio continuum emission, see Section~4.3).
   Two \sfrghz\ estimates are given: (1) for $M > 5$\Msun\ (recent SF) and 
   (2) for $M > 0.1$\Msun\ (overall SF, labeled $M_{\star}$).}
\end{figure}

\begin{table} 
\caption{Luminosities and star formation rates [\MMoy]}
\label{tab:sfr} 
\begin{tabular}{lcccc}
\hline
                     & NGC~1512          & NGC~1510           \\
\hline
$L_{\rm FUV}$ [$10^{38}$ erg\,s$^{-1}$ \AA$^{-1}$]
	             & $14.2 \pm 0.7$    & $4.7 \pm 0.3$      \\
$L_{\rm H\alpha}$ [10$^{40}$ erg\,s$^{-1}$]        
                     & $2.4 \pm 1.1$     & $0.9 \pm 0.1$      \\  
$L_{\rm 24\mu m}$ [10$^{41}$ erg\,s$^{-1}$]  
		     & $6.3 \pm 0.3$     & $2.0 \pm 0.2$(*)   \\
$L_{\rm 60\mu m}$ [10$^{22}$ W\,Hz$^{-1}$]  
	             & $3.37 \pm 0.15$   & $0.97 \pm 0.10$    \\
$L_{\rm FIR}$ [10$^{41}$ erg\,s$^{-1}$]             
                     & $26.0 \pm 1.4$    & $4.6 \pm 0.4$      \\
$L_{\rm 20cm}$ [10$^{19}$ W\,Hz$^{-1}$]  
                    & $35.9 \pm 0.6$  & $4.2 \pm 0.6$         \\
\hline
\sfrfuv\ (S07)      & $0.115 \pm 0.006$   & $0.038 \pm 0.002$ \\ 
\sfrfuv\ (K98)      & $0.150 \pm 0.008$   & $0.049 \pm 0.003$ \\ 
\sfrha\ (C07)       & $0.13~ \pm 0.06$    & $0.047 \pm 0.007$ \\
\sfrha\ (K98)       & $0.19~ \pm 0.09$    & $0.070 \pm 0.010$ \\
$SFR_{24\mu m}$     & $0.075 \pm 0.004$   & $0.041 \pm 0.005$ \\
\sfrmir\            & $0.22~ \pm 0.08$    & $0.072 \pm 0.014$ \\            
$SFR_{60\mu m}$     & $0.066 \pm 0.003$   & $0.019 \pm 0.002$ \\
\sfrfir\            & $0.117 \pm 0.006$   & $0.021 \pm 0.002$ \\
\sfrghz ($\rm M\ge 5 M_{\odot}$)  
                    & $0.105 \pm 0.002$   & $0.012 \pm 0.002$ \\
\sfrghz ($\rm M\ge 0.1 M_{\odot}$)
                    & $0.500 \pm 0.008$   & $0.058 \pm 0.008$ \\
\hline
\end{tabular}
\flushleft
References: 
   Gil de Paz et al. (2007a, $FUV$), 
     uncertainties in the $FUV$ flux were estimated in here; 
   Meurer et al. (2006, \Ha);
   Dale et al. (2007, Spitzer 24$\mu$m) for NGC~1512, 
     (*) $L_{\rm 24\mu m}$ for NGC~1510 was estimated in here; 
   Moshir et al. (1990, IRAS $FIR$);
   K98 = Kennicutt (1998), 
   S07 = Salim et al. (2007), 
   C07 = Calzetti et al. (2007). 
\end{table}

\subsection{The global star formation rate} 

There are numerous ways to estimate the star formation rate (SFR) of a galaxy.
To study the global and local SFRs, we use a range of line and continuum 
measurements at different wavelengths (ultraviolet, optical, infrared, and 
radio). A combination of these data together with an understanding of which
stellar populations are detected at each wavelength is essential to obtain 
the full picture. Nevertheless, we are somewhat limited by the sensitivity, 
quality and field-of-view of the existing observations. 

Star formation tends to be localised and varies within galaxies. While the 
nuclear region and inner spiral arms of a galaxy are generally locations of 
significant star formation, we also find new stars forming in other areas such 
as interaction zones and occasionally in isolated clumps (presumably of high 
molecular gas density) in the far outskirts of galaxies. The NGC~1512/1510 
system is an excellent laboratory to study the locations and properties of 
its many star forming regions, from the galaxy nuclei out to the largest radii 
where detached \HI\ clouds are found (see Section~3.3) as well as in the 
interaction zone between the two galaxies. \\

Here we use a range of tracers to study the global SFR of both NGC~1512 and 
NGC~1510 (results are summarised in Table~\ref{tab:sfr} and Fig.~14), before 
investigating the local star formation activity within various parts of the 
NGC~1512/1510 system (see Section~4.4).

From our 20-cm radio continuum data we derive a recent global SFR of 
0.105\Moy\ for NGC~1512 and 0.012\Moy\ for NGC~1510 (see Section~3.4). 
Another extinction-free SFR estimate is derived from the far-infrared ($FIR$) 
luminosity. Using the IRAS flux densities (Moshir et al. 1990) together with
the relations given by Sanders \& Mirabel (1996) and Kennicutt (1998), we 
derive \sfrfir\ $\approx$ 0.12\Moy\ for NGC~1512 and 0.02\Moy\ for NGC~1510.

$FIR$ emission comes from the thermal continuum re-radiation of dust grains 
which absorb the visible and $UV$ radiation emitted by massive young stars. In 
contrast, radio continuum emission is mainly due to synchrotron radiation from 
relativistic electrons accelerated in the remnants of core-collapse supernovae,
therefore also associated with the presence of massive stars. Both estimates 
trace the star formation activity in the last $\sim$100~Myr. However, as 
relativistic electrons have lifetimes of $\sim$100~Myr (Condon et al. 2002), 
we should expect that the 20-cm radio continuum emission traces SFRs with 
somewhat extended ages.

\Ha\ emission traces the most massive, ionising stars, and timescales of
$\sim$10~Myr, i.e. the most recent events of star formation in the galaxy.
The \Ha\ flux given by Meurer et al. (2006) was corrected for Galactic 
extinction but not for internal extinction or for the contribution of the 
\NII\ emission lines adjacent to \Ha\ (see L\'opez-S\'anchez \& Esteban 
2008)\footnote{As NGC~1510 is a low-metallicity galaxy, the contribution of 
  the \NII\ emission to the \Ha\ flux is expected to be negligible.}.
Using the relation by Kennicutt (1998), we find \sfrha\ = 0.19 and 0.07\Moy\ 
for NGC~1512 and NGC~1510, respectively. Slighter lower values, \sfrha\ = 
0.13 and 0.05\Moy, result when using the more recent Calzetti et al. (2007) 
calibration. 

$UV$-emission probes star formation over timescales of $\sim$100~Myr, the 
life-time of the massive OB stars. Using the extinction-corrected 
GALEX $UV$-magnitude, $m_{\rm FUV}$, as given by Gil de Paz et al. (2007a), 
we derive the $UV$-flux as follows: 
   $f_{\rm FUV}$ [erg\,s$^{-1}$\,cm$^{-2}$\,\AA$^{-1}$] = 
   $1.40 \times 10^{-15} \times 10^{0.4 \times (18.82 - m_{\rm FUV})}$.
We have corrected $m_{\rm FUV}$ for extinction assuming the Galactic value 
provided by Schlegel et al. (1998), $E(B-V)$ = 0.011, and $A_{\rm FUV} = 
7.9~E(B-V)$. Applying the Salim et al. (2007) relation between the $FUV$ 
luminosity and the SFR, we obtain \sfrfuv\ = 0.12 and 0.04\Moy\ for NGC~1512 
and NGC~1510, respectively. For comparison, applying the Kennicutt (1998) 
relation results in values that are a 1.3 times higher. Here we prefer to 
use Salim et al. (2007) relation because it was derived using GALEX data.

The SINGS Legacy project (Kennicutt et al. 2003) provides Spitzer mid-infrared 
($MIR$) images of NGC~1512/1510. $MIR$ emission, which traces the dust 
distribution within galaxies, also agrees well with the position of the 
$UV$-rich star clusters in the system. Because of its higher intrinsic 
brightness, the $MIR$ emission is mainly detected in the cores of both 
galaxies and in the inner ring of NGC~1512. Using the Spitzer 24$\mu$m flux 
density measurements of NGC~1512 (Dale et al. 2007) and NGC~1510 (obtained by 
us; see Table~5) together with the relations by Calzetti et al. (2007) we 
derive $SFR_{24\mu m}$ = 0.075\Moy\ for NGC~1512 and 0.041\Moy\ for NGC~1510.

Combining the 24$\mu$m luminosity (which traces the dust-absorbed star 
formation) with the \Ha\ luminosity (which probes the unobscured star 
formation) we derive $SFR_{\rm H\alpha+24\mu m}$ = 0.22\Moy\ and 0.07\Moy\ 
for NGC~1512 and NGC~1510, respectively. \\

Figure~14 shows the star formation rates as derived for NGC~1512 and NGC~1510 
at various wavelength, arranged along the x-axis by the approximate timescales 
in which the SFR is considered: from the \Ha\ emission, tracing the very young 
($\sim$10~Myr) star formation to the radio continuum emission, tracing the old 
stellar population ($\sim$100~Myr). We find that for NGC~1510 our derived SFR 
estimates are in agreement ($\sim$0.05\Moy). The fact that the SFR derived 
from the 20-cm radio continuum flux considering all masses, \sfrghz\ ($M \ge 
0.1$\Msun) = 0.058\Moy, is close to the SFR found using $FUV$, \Ha\ and 
24$\mu$m data reinforces the starburst nature of this BCD galaxy. 
For NGC~1512, the SFR estimates obtained from \Ha, $FUV$, 24$\mu$m, $FIR$ and 
20-cm radio continuum ($M > 5$\Msun) data agree ($\sim$0.12\Moy). However, 
\sfrghz\ is about four times higher when we consider all masses. This can be 
explained by the non-starbursting nature of NGC~1512, which has been forming 
stars over a long period of time ($\sim$Gyr).  \\

Following Helou, Soifer \& Rowan-Robinson (1985) we calculate the $q$ parameter
which is defined as the logarithmic ratio of the $FIR$ to 20-cm radio flux 
density. We find $q$ = 2.25 and 2.44 for NGC~1512 and NGC~1510, respectively, 
consistent with the mean value of 2.3 for normal spiral galaxies (Condon 1992).
This result confirms the star-forming nature of both galaxies. \\

Using the \Ha\ flux given by Meurer et al. (2006) and the relation provided 
by Condon et al. (2002), we derive the thermal flux at 1.4 GHz for NGC~1512 
and NGC~1510: 2.7 and 1.0 mJy, respectively. The ratio of the non-thermal to 
thermal radio emission, $\log R$, is 1.1 and 0.54, respectively. The value 
derived for NGC~1512 agrees with that of typical star-forming galaxies ($\log 
R = 1.3 \pm 0.4$, Dopita et al. 2002) but is relatively low for NGC~1510. 
This indicates that the thermal emission from \HII\ regions in NGC~1510 is 
more important than the non-thermal emission from supernovae explosions (i.e., 
the starburst is very recent, and there has not been enough time to convert 
many massive stars into supernovae). This fact agrees with the detection of 
WR features in NGC~1510 (Eichendorf \& Nieto 1984).

\begin{figure} 
\label{fig:sfrha-uv}
  \mbox{\psfig{file=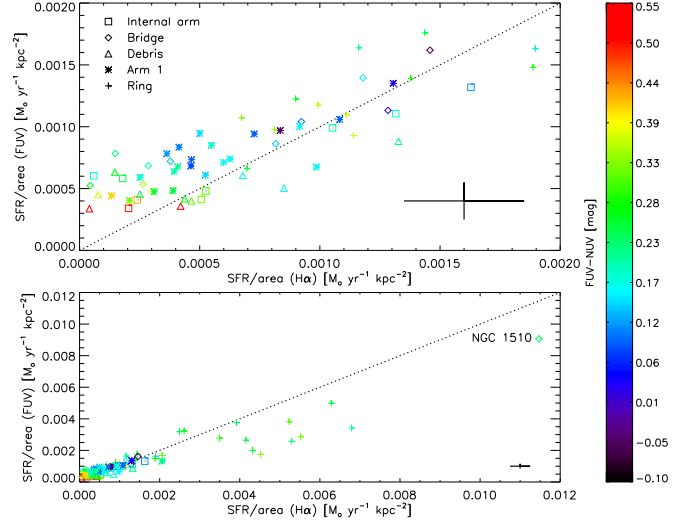,width=7cm,angle=-270}}
\caption{Comparison of \sfrha/area (x-axis) with \sfrfuv/area (y-axis) for
   stellar clusters detected in both \Ha\ and $UV$ emission. Different symbols 
   indicate distinct regions within the NGC~1512/1510 system. The $FUV-NUV$ 
   color scale ranges between --0.1 (black) and +0.55 (red). The top diagram 
   is an enlargement of the bottom one. The dotted line indicates the place 
   where \sfrha/area = \sfrfuv/area. Typical error bars are shown at the right 
   side of each diagram.}
\end{figure}

\begin{figure} 
\label{fig:sfrha-mir}
  \mbox{\psfig{file=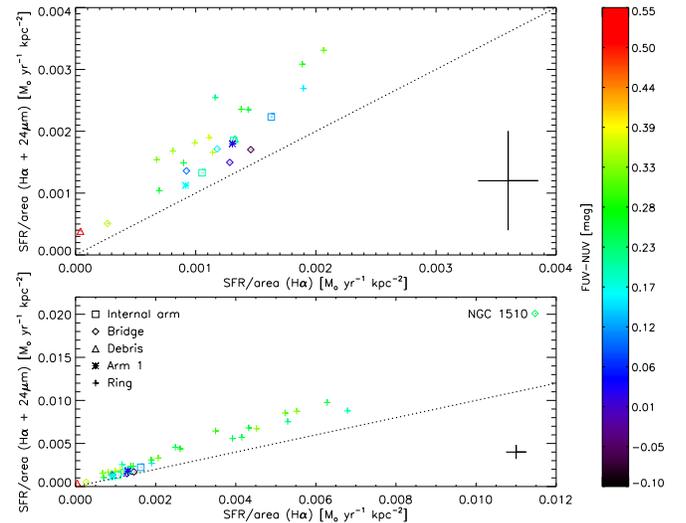,width=7cm,angle=-270}}
\caption{Comparison of \sfrha/area (x-axis) with \sfrmir/area (y-axis) for 
   stellar clusters detected in both \Ha\ and 24$\mu$m emission. Different 
   symbols indicate distinct regions within the NGC~1512/1510 system. The 
   $FUV-NUV$ color scale ranges between --0.1 (black) and +0.55 (red). The 
   top diagram is an enlargement of the bottom one. The dotted line indicates 
   the place where \sfrha/area = \sfrmir/area. Typical error bars are shown 
   at the right side of each diagram.}
\end{figure}

\subsection{The local star formation activity} 

We have estimated the star formation rate of each $UV$-rich stellar cluster in
the NGC~1512/1510 system using the extinction-corrected $FUV$ luminosity and 
the assumptions given before. In general, we find that regions closer to 
NGC~1512 display higher star formation activity, in agreement with their young
ages (see Fig.~13). 

Considering only stellar clusters with both \Ha\ and $UV$ emission, we compare
\sfrha\ (C07) with \sfrfuv\ (S07). Fig.~15 shows a good correlation between 
both estimates for small values of SFR/area ($\leq$0.002\Moy\,kpc$^{-2}$), 
however, \sfrfuv\ is always lower than \sfrha\ for regions with high SFR/area
(i.e. within the inner ring of NGC~1512 and in NGC~1510). Given that the ages
of the latter are similar, this must be a consequence of internal extinction 
within these regions which are denser and possess a larger amount of dust, as 
seen in Spitzer images, than other areas. Fig.~16 confirms this as \sfrmir\ is 
systematically higher than \sfrha.

Next, we investigate if the $UV$-rich clusters within the NGC~1512/1510 system 
do obey the Schmidt-Kennicutt scaling laws of star formation (Kennicutt 1998).
Boissier et al. (2007), for example, find that the stellar and gas radial 
profiles of galaxies with $XUV$ disks follow such relations. Fig.~17 shows a 
comparison between \sfrfuv/area and the \HI\ mass density. --- This analysis 
can be improved by adding high-resolution ATCA \HI\ data, obtained for the 
southern THINGS project (Deanne, de Blok, et al.), to improve the sensitivity 
to small scale structure, 

\begin{figure} 
\label{fig:sfruv} 
  \mbox{\psfig{file=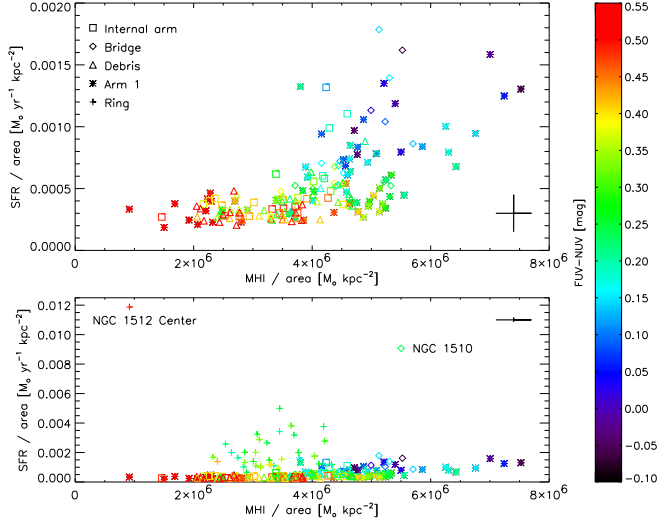,width=7cm,angle=-270}} 
\caption{Comparison of \sfrfuv/area with the \HI\ mass density for all analysed 
   $UV$-rich clusters. Different symbols indicate distinct regions within the 
   NGC~1512/1510 system. The $FUV-NUV$ color scale ranges between --0.1 (black)
   and +0.55 (red). The top diagram is an enlargement of the bottom one, but
   omitting $UV$-regions within the inner ring of NGC~1512 (+ symbols). Typical
   error bars are shown at the right side of each diagram.}
\end{figure}

\begin{figure} 
\label{fig:density} 
  \mbox{\psfig{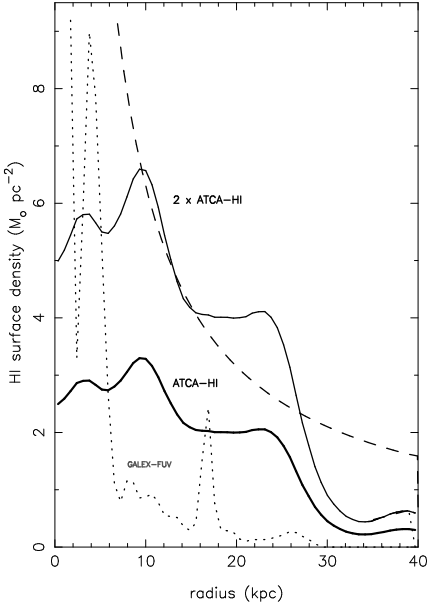}} 
\caption{Radially integrated \HI\ surface density of NGC~1512 (solid line) 
   compared to the critical density (dashed line) estimated for $\alpha_{\rm Q}
   = 0.7$ and \vrot\ = 150\kms. We also show the radial density of the GALEX 
   FUV emission (dotted line) which is an excellent tracer of star formation 
   activity. Due to the lack of CO data we cannot show the total gas density, 
   which is expected to be close or above the critical density at all radii 
   where star formation is present.}
\end{figure}

\begin{figure} 
\label{fig:histo}  
   \mbox{\psfig{file=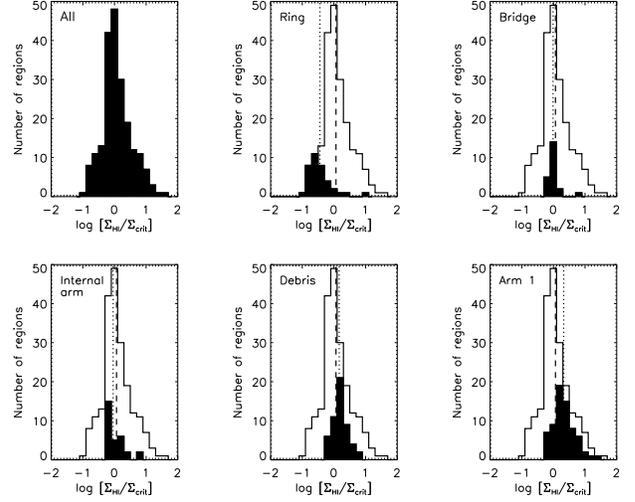,width=7cm,angle=-270}}
\caption{Histograms of the logarithmic ratio of the \HI\ gas density, 
   $\Sigma_{\rm HI}$, and the critical density, $\Sigma_{\rm crit}$, for 
   $UV$-rich clusters in the NGC~1512/1510 system. Each panel shows a 
   distinct region (filled area), apart from the top left panel which 
   shows all analysed stellar clusters. For comparison, the latter 
   is also overlaid onto the histograms of the other regions. The dashed 
   vertical line indicates the average ratio for all clusters, while the 
   dotted line indicates the average ratio for each region.}
\end{figure}

Overall, we find that \sfrfuv/area increases with \MHI/area, i.e. the higher 
the \HI\ gas density the more stars are forming. However, this is not true for 
regions located within the inner star-forming ring of NGC~1512, where there is
clearly a lack of \HI\ gas (see Fig.~13). We conclude that in the inner region 
of NGC~1512 a large amount of molecular gas must be present to boost the 
overall gas density to the critial value or above. \\

In the following we check if the Toomre Q gravitational stability criterion is 
satisfied at the locations of the $UV$-rich clusters. Ideally, we would use 
the \HI\ velocity field (corrected for inclination) and the velocity dispersion 
of NGC~1512, to compute the critical gas density, 
   $\Sigma_{\rm crit} = \alpha_{\rm Q} \frac{\sigma \kappa}{\pi G}$ (see
Kennicutt 1989, Martin \& Kennicutt 2001), at every pixel in the disk. Here 
$\alpha_{\rm Q}$ is a scaling constant, $\sigma$ is the velocity 
dispersion, and $\kappa$ is the epicyclic frequency. The low-resolution \HI\ 
distribution (0.\,moment) and the mean \HI\ velocity field (1.\,moment) of 
NGC~1512 are shown in Fig.~3. The \HI\ velocity dispersion (2.\,moment, not 
shown) varies between 7--22\kms\ in the spiral/tidal arms of NGC~1512. As a
first step, we compare the radially averaged \HI\ gas distribution with the
critical density and the radially averaged $FUV$ emission. For a flat rotation 
curve (i.e., \vrot($r$) = constant), which is a reasonable assumption for $r
\ga 15\arcsec$ (see Fig.~6), and a velocity dispersion of 6\kms\ (as used in 
previous work), the above equation reduces to 
   $\Sigma_{\rm crit}$ = 0.6 $\alpha_{\rm Q}$ \vrot/$r$, where $r$ is the 
radius in kpc. Using the derived \HI\ inclination ($i$ = 35\degr) and position 
angle ($PA$ = 265\degr), the de-projected \HI\ radial surface density, 
$\Sigma_{\rm HI}$ of NGC~1512 is shown in Fig.~18. The critcal density was 
computed for $\alpha_{\rm Q}$ = 0.7 and \vrot\ = 150\kms. We find that the 
radially averaged \HI\ gas alone lies just below the computed critical density.

At radii less than 10~kpc, an increasing amount of molecular gas is needed to 
reach critical gas density and feed the star formation in the nuclear region
and the inner ring. The radially integrated $FUV$ flux drops below the noise 
at $r$ = 28 kpc, the likely SF threshold.

In the inner region of NGC~1512, the gas motions are strongly affected by the 
stellar bar, which would affect the critical density estimate.
Using NGC~1512's angular velocity, $\Omega$($r$) = \vrot($r$)/$r$, we can also 
determine the locations of the inner and outer Lindblad resonances: 
  $\Omega_{\rm p} = \Omega(r) \pm \kappa(r)/2$, where $\Omega_{\rm p}$ is the 
bar pattern speed and $\kappa(r) = \sqrt2$ \vrot/$r$. At $r \approx 4$ kpc
(bar radius) its pattern speed is $\sim$50\kms\,kpc$^{-1}$, suggesting that 
the ILR(s) lie at $r \la 1.2$ kpc, and the OLR at $r \sim 6.8$ kpc. \\

Figure~19 shows the logarithmic ratio of the measured \HI\ gas density,
$\Sigma_{\rm HI}$, to the critical density, $\Sigma_{\rm crit}$, for all 
analysed $UV$-rich clusters together and within the previously defined 
distinct regions. Taking all clusters, a peak is found at 
$<\log(\Sigma_{\rm HI}/\Sigma_{\rm crit})> = 0.06$, indicating that local star 
formation is, on average, happening at the local critical density. The measured
\HI\ densities are slightly higher than critical along Arm~1 (0.33) and within 
the NW debris (0.17), but significantly lower than critical in the inner 
star-forming ring (--0.44) where the molecular gas density must be high. 

Finally, Fig.~20 shows -- on a logarithmic scale -- the \sfrfuv\ density 
versus the gas density for $UV$-rich clusters in the NGC~1512/1510 system 
(derived here) and in the nearby Sbc galaxy M\,51 (Kennicutt et al. 2007;
for $r \la 8$ kpc). 
Because no molecular data are currently available for NGC~1512, only the \HI\ 
gas density is shown. Regions within the inner star-forming ring of NGC~1512 
(located at $r$ = 90\arcsec) are -- as stated before -- significantly offset.
Tripling the \HI\ mass of each region achieves an approximate alignement. For 
M\,51, molecular data are taken into account and are found to be essential for 
the observed correlation (Kennicutt et al. 2007). 
Dong et al. (2008) found that the $UV$-selected regions in two small fields
within the large gaseous disk of M\,83, $\sim$20 kpc from its centre, follow 
a similar trend. Molecular gas was not taken into account (for comparison,
see Martin \& Kennicutt 2001).

\begin{figure} 
\label{fig:kennicutt} 
  \mbox{\psfig{file=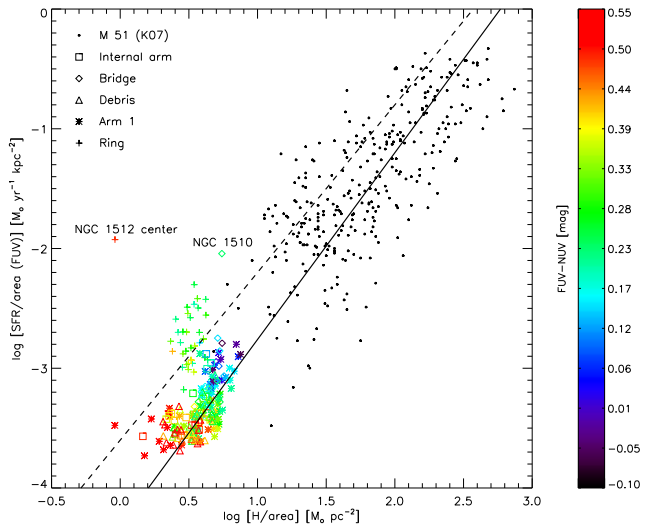,width=7cm,angle=-270}} 
\caption{Relation between \sfrfuv/area and the \HI\ gas density for $UV$-rich 
   clusters in the NGC~1512/1510 system (derived here) and the galaxy M\,51 
   (from Kennicutt et al. 2007). As before, different symbols indicate distinct
   regions within the NGC~1512/1510 system. The $FUV-NUV$ colors range from 
   --0.1 (black) to +0.55 (red). The solid line is the best fit to the M\,51 
   data (Kennicutt et al. 2007); the dashed line is the relation for whole 
   galaxies derived by Kennicutt (1998). The gas density for M\,51 was derived 
   combining atomic and molecular gas measurements, whereas we only use the 
   \HI\ gas for NGC~1512. }
\end{figure}

\subsection{Global chemical properties} 

Here we estimate the metallicities of NGC~1512 and NGC~1510 using data 
available in the literature. Calzetti et al. (2007) and Moustakas \& 
Kennicutt (2006) give an oxygen abundance between 8.37 and 8.81, in 
units of 12+log(O/H), for the chemical abundance of NGC~1512. The first 
value was derived using optical spectroscopy and the Pilyugin \& Thuan (2005) 
calibration. The second value was obtained comparing with the predictions 
given by the photoionised evolutionary synthesis models provided by 
Kobulnicky \& Kewley (2004). However, some recent analysis using direct 
estimates of the electron temperature ($T_e$) of the ionised gas (e.g., 
L\'opez-S\'anchez 2006) suggest that these models overestimate the oxygen 
abundance by $\sim$0.2 dex. We conclude that the metallicity of NGC~1512 is 
between 8.4 and 8.6, slightly lower than for the Milky Way, but within the 
range typical observed for spiral galaxies (Henry \& Worthey 1999).

On the other hand, we have used the emission line intensity data for NGC~1510, 
provided by Storchi-Bergmann et al. (1995), to compute its chemical abundance. 
We used the \Ha\/\Hb\ ratio to correct the data for reddening and obtained 
$C$(\Hb) = 0.54. With the help of tasks in the {\sc iraf} `nebular package', 
we compute $T_{\rm e}$ = 15700 K, using the [O\,{\sc iii}] 
$\lambda$5007/$\lambda$4363 ratio. Assuming an electron density of $n_{\rm e}$ 
= 100~cm$^{-3}$ we compute the ionic abundances for O$^+$/H$^+$ and 
O$^{++}$/H$^+$ and derive a total oxygen abundance of 12+log(O/H) = 7.95 
($\sim$0.2\Zsun), typical for BCD galaxies. 

The large metallicity difference between NGC~1512 and NGC~1510 indicates that 
both galaxies have experienced a very different chemical evolution, and that
NGC~1510 has been in a quiescent state for a long time while NGC~1512 was 
forming stars continuously.  

The N/O ratio found in NGC~1510, log(N/O) $\approx$ --1.2, is rather high for 
a galaxy with its oxygen abundance. For comparison, Izotov \& Thuan (2004) 
typically obtain log(N/O) $\approx$ --1.5 for low metallicity BCD galaxies. 
Similar results are also found for other BCD galaxies with a significant 
population of WR stars (Brinchmann et al. 2008). It is thought that the 
nitrogen enrichment is a consequence of a very recent chemical pollution 
event probably connected with the onset of WR winds (L\'opez-S\'anchez et al. 
2007). It may also be related to the interaction between galaxies (Pustilnik 
et al. 2004, L\'opez-S\'anchez et al. 2009), but deeper optical spectroscopic 
data with a higher spectral resolution are needed to confirm this issue.

\subsection{Interaction-induced star formation}  

Star formation depends on the gravitational instability of galaxy disks,
both locally and globally. Minor mergers and tidal interactions affect
the gas distribution and dynamics of galaxies, leading to the formation of 
bars, gas inflow as well as the ejection of gas, and -- as a consequence 
-- locally enhanced star formation. Together, these phenomena are key 
ingredients to the understanding of galaxy evolution. The often extended
\HI\ envelopes of spiral galaxies are particularly useful as sensitive
tracers of tidal interactions and gas accretion. The gas distribution and 
dynamics are easily influenced by the environment, resulting in asymmetries, 
line broadening and/or splitting etc. 

The development of a strong two-armed spiral pattern and star-forming regions
in disk galaxies (here NGC~1512) which accrete low-mass dwarf companions (here 
NGC~1510) has been explored by Mihos \& Hernquist (1994) using numerical 
simulations. Their models, which use a mass ratio of 10:1 for the disk galaxy 
and its companion, resemble the galaxy pair NGC~1512/1510 after $\sim$40 time 
units (i.e. $5.2 \times 10^8$ years). At that stage, the model disk galaxy
has developed a pronounced, slightly asymmetric two-armed spiral pattern with
significant star-formation along the arms and the nuclear region.

Minor mergers are common. The Milky Way and the Andromeda galaxy are prominent
examples; both have many satellites and show evidence for continuous accretion 
of small companions. The multitude of stellar streams detected in our Galaxy
as well as some other galaxies (e.g., NGC~5907, Martinez-Delgado et al. 2008) 
are hinting at a rich accretion history. Minor mergers contribute significantly
to galaxy assembly, accretion, and evolution.

\section{Conclusions}

We analysed the distribution and kinematics of the \HI\ gas as well as the 
star formation activity in the galaxy pair NGC~1512/1510 and its surroundings. 

For the barred, double-ring galaxy NGC~1512 we find a very large \HI\ disk,
about four times the optical diameter, with two pronounced spiral arms, 
possibly tidally induced by the interaction with the neighbouring blue compact 
dwarf galaxy NGC~1510. It is possible that the interaction also triggered
the formation of the bar in NGC~1512 (unless the bar already existed, maybe
from a previous accretion or interaction event) which would then cause gas to 
fall towards the nuclear regions, feeding the star formation, as well as
induces torques in the outer spiral arms.

We detect two {\em tidal dwarf galaxies} with \HI\ masses of $\la10^7$\Msun\
and clear signs of star formation in the outer-most regions of the system. The 
most distant TDG, N1512-west, is rather compact and lies at a distance of 
$\sim$80 kpc from the centre of NGC~1512, potentially at the tip of an extra\-
polated eastern \HI\ arm of NGC~1512. The second TDG, 1512-south, is forming 
within an extended \HI\ cloud, and is located slightly closer (64 kpc), within 
the extrapolated eastern \HI\ arm. 

We regard these two TDGs as typical with respect to their \HI\ mass, star 
forming activity and detachment from the interacting system. While TDGs are 
often found in major mergers, we find that they can form in mildly interacting 
system such as NGC~1512/1510. In this case, the interaction is effectly an 
accretion of a blue compact dwarf galaxy (NGC~1510) by the large spiral galaxy 
NGC~1512. 

NGC~1512 hosts an extended $UV$ disk with $\ga$200 of clusters with recent
star formation activity. The comparison of our \HI\ map with the GALEX 
images clearly shows that these clumps are located within the maxima
of neutral gas density. We have derived the ages and star formation rates 
of the $UV$-rich clusters. 

We find that generally only the youngest $UV$ clusters are associated with
high \HI\ column densities, while in older $UV$ clusters only diffuse \HI\ gas
is detected. This might suggest that as the hydrogen gas depleted, star 
formation stopped in the latter regions. As a consequence we expect to detect
\Ha\ emission in all high density \HI\ regions or equivalent in all young 
$UV$ clusters.

Our analysis supports a scenario in which the interaction between the BCD 
galaxy NGC~1510 and the large spiral galaxy NGC~1512 has triggered star 
formation activity in the outskirts of the disk and enhanced the tidal
distortion in the \HI\ arms. The interaction seems to occur in the north
western areas of the system because of the broadening of the \HI\ arm
and the spread of the $UV$-rich star clusters in this region.
The system is probably in the first stages of a minor merger which started 
$\sim$400~Myr ago. 

\section{Outlook}

Future \HI\ surveys, such as those planned with the Australian SKA Pathfinder
(ASKAP; Johnston et al. 2008) will produce similar \HI\ cubes and images than 
obtained here, but over much larger areas. E.g., the proposed shallow ASKAP 
\HI\ survey of the sky will reach a sensitivity of $\sim$1 mJy\,beam$^{-1}$ at 
an angular resolution of 30\arcsec\ in a 12-h integration per field. Focal 
plane arrays will provide a very large, instantaneous field-of-view of $5\fdg5 
\times 5\fdg5$. This means that \HI\ images similar to those shown in this 
paper will be obtained for the entire Local Volume. Furthermore, the correlator
bandwidth of 300 MHz (divided into 16,000 channels) will allow us to study the
\HI\ content of galaxies and their surroundings out to $\sim$60,000\kms\ ($z$ 
= 0.2). In addition, very deep 20-cm radio continuum images are obtained 
for the same area.

\section*{Acknowledgements}
\begin{itemize}
\item This research has made extensive use of the NASA/IPAC Extragalactic 
      Database (NED) which is operated by the Jet Propulsion Laboratory, 
      Caltech, under contract with the National Aeronautics and Space 
      Administration. 
\item The Digitised Sky Survey was produced by the Space Telescope Science
      Institute (STScI) and is based on photographic data from the UK Schmidt 
      Telescope, the Royal Observatory Edinburgh, the UK Science and 
      Engineering Research Council, and the Anglo-Australian Observatory.
\item We thank David Malin for permission to use the deep optical image of 
      the NGC1512/1510 system.
\end{itemize}

\end{document}